\documentclass[english,onecolumn,12pt,draftclsnofoot]{IEEEtran}
\usepackage[T1]{fontenc}
\usepackage[latin9]{inputenc}
\usepackage{verbatim}
\usepackage{float}
\usepackage{mathtools}
\usepackage{amsmath}
\usepackage{amsthm}
\usepackage{amssymb}
\usepackage{graphicx}

\makeatletter

\floatstyle{ruled}
\newfloat{algorithm}{tbp}{loa}
\providecommand{\algorithmname}{Algorithm}
\floatname{algorithm}{\protect\algorithmname}

\theoremstyle{plain}
\ifx\thechapter\undefined
	\newtheorem{thm}{\protect\theoremname}
\else
	\newtheorem{thm}{\protect\theoremname}[chapter]
\fi
\theoremstyle{plain}
\newtheorem{cor}[thm]{\protect\corollaryname}

\usepackage{cite}
\usepackage{tikz,pgf}
\usetikzlibrary{calc,arrows}
\usepackage{algorithm,algpseudocode}

\makeatother

\usepackage{babel}
\providecommand{\corollaryname}{Corollary}
\providecommand{\theoremname}{Theorem}

\begin{document}
\title{Online Learning Models for Content Popularity Prediction In Wireless
Edge Caching}
\author{Navneet Garg, \IEEEmembership{Student Member, IEEE}, Vimal Bhatia,
\IEEEmembership{Senior Member, IEEE}, \\B. N. Bharath, \IEEEmembership{Member, IEEE},
Mathini Sellathurai, \IEEEmembership{Senior Member, IEEE}, Tharmalingam
Ratnarajah, \IEEEmembership{Senior Member, IEEE} }
\maketitle
\begin{abstract}
Caching popular contents in advance is an important technique to achieve
the low latency requirement and to reduce the backhaul costs in future
wireless communications. Considering a network with base stations
distributed as a Poisson point process (PPP), optimal content placement
caching probabilities are derived for known popularity profile, which
is unknown in practice. In this paper, online prediction (OP) and
online learning (OL) methods are presented based on popularity prediction
model (PPM) and Grassmannian prediction model (GPM), to predict the
content profile for future time slots for time-varying popularities.
In OP, the problem of finding the coefficients is modeled as a constrained
non-negative least squares (NNLS) problem which is solved with a modified
NNLS algorithm. In addition, these two models are compared with log-request
prediction model (RPM), information prediction model (IPM) and average
success probability (ASP) based model. Next, in OL methods for the
time-varying case, the cumulative mean squared error (MSE) is minimized
and the MSE regret is analyzed for each of the models. Moreover, for
quasi-time varying case where the popularity changes block-wise, KWIK
(know what it knows) learning method is modified for these models
to improve the prediction MSE and ASP performance. Simulation results
show that for OP, PPM and GPM provides the best ASP among these models,
concluding that minimum mean squared error based models do not necessarily
result in optimal ASP. OL based models yield approximately similar
ASP and MSE, while for quasi-time varying case, KWIK methods provide
 better performance, which has been verified with MovieLens dataset. 
\end{abstract}


\section{Introduction \label{sec:Introduction}}

With the continuous development of various intelligent devices and
application services, wireless mobile communications has been experiencing
an unprecedented traffic surge with a lot of redundant and repeated
information \cite{8531745}. This redundant traffic limits the fronthaul
and backhaul capacity. In the recent years, edge caching has been
used as an effective method for reducing peak data rates by pre-storing
the most popular contents in the local cache of a base station \cite{shanmugam2013femtocaching}.
In edge caching, the traffic load during peak periods is shifted to
off-peak periods, by fetching the \textquotedblleft anticipated\textquotedblright{}
popular contents, e.g., reusable video streams are stored in local
cache, and are reused during off-peak hours \cite{poularakis2016complexity}.
Accordingly, in \cite{blaszczyszyn2015optimal,serbetci2017optimal},
content placement in cellular networks is optimized to maximize cache
hit rate, while \cite{liu2017caching,liu2016caching} obtain optimal
placement policy to maximize the success probability and area spectral
efficiency. On a similar note, \cite{avrachenkov2017low,avrachenkov2017optimization}
relies on minimizing cache miss probability to get caching policy.
In \cite{maddah2014fundamental}, lower bounds for local and global
caching are derived. In \cite{sadeghi2018optimal}, Q-learning is
employed to obtain the caching policy when the popularity profile
in different time slots is modeled as a Markov chain. These works
shows that in a given network, caching policy can be obtained when
the popularity profile is known in advance. However, in practice,
popularity profile is not known and requires to be estimated from
the past observations of content requests. Several works employ different
models for forecasting popularity profiles. \cite{yin2018prediction,liu2018content}
employs neural networks and deep learning based approaches for prediction.
The drawback of this approach is the requirement of huge data and
time for training. \cite{nakayama2015caching} models popularity using
auto regressive (AR) model to predict the number of requests in the
time series. \cite{zhang2018ppc} predicts content requests for video
segments using a linear model. 
In \cite{7775114}, a context-aware online policy has
been presented which learns content popularity independently across
contents. The authors in \cite{7524790} have proposed a low complexity online policy
for video caching which assumes that the expected popularities of
similar contents are similar. These works assume the content popularity
remains unchanged for a certain time period. In practical scenarios,
the content popularity changes dynamically in both time and space
dimensions owing to randomness of user requests. Also, in most of
the works, the content placement with respect to change in popularity
is not optimized. In this regard, for different prediction models,
we investigate that minimum mean squared error (MSE) based models
do not necessarily result in optimal performance metric. Moreover,
although the content popularity depends on context-based parameters
including location, time, the type of file, etc., the popularity prediction
can be modeled using a linear model of past popularities, as investigated
in most of the literature. The noticeable fact is that the parameters/variables
controlling the popularity are already present in the past observations.
The goal is to find a smart prediction method for a given network
in order to provide enhanced quality of service to users with the
given cache objective, allowing service differentiation requirement.

In this paper, we present online prediction as well as online learning
models to predict the content probabilities considering both time-varying
scenario, and KWIK based models for quasi-time-varying (block-wise)
scenarios. First, assuming a network where both the BS and users are
distributed as homogeneous PPP and content requests are characterized
using a global popularity profile, we find optimum content placement
caching probabilities to maximize the ASP, when the profile is known.
Next, introducing time varying scenario where content distribution
changes independently in each time slot, popularity profiles, are
predicted in an online fashion via five models, in which first two
(popularity prediction model (PPM) and Grassmannian prediction model
(GPM)) are novel ones to the best of authors\textquoteright{} knowledge,
while the remaining three (namely information prediction model (IPM),
request prediction model(RPM), and ASP based prediction model (ASP-PM))
have been employed in literature in some different ways. In the first
two models (PPM and GPM), the problem of finding the coefficients
of linear regression is formulated as a constrained non-negative least
squares (NNLS) optimization, for which a modified NNLS algorithm is
provided. Moreover, in time varying case, based on PPM and GPM, online
learning approaches have been presented in order to improve the solution
from online predictions. Independence of popularities in the time
varying case is a strong assumption, since predicting the popularity
profile which is correlated across time slots, is easier than the
case when it is independent. In most of the works, correlation based
time-varying models are considered, which is called here quasi-time
varying case. In quasi-time-varying case, we assume the non-negative
coefficients of linear regression are constant in a block and are
changed in the next block. For a given block, it is like a convex-auto-regressive
process. For this scenario, online learning methods based on KWIK
(\textquotedblleft know what it knows\textquotedblright ) framework
are modified for prediction. In simulations, prediction MSE and ASP
difference are compared which show that for time varying case, GPM
and PPM based online learning methods provide better performance,
while for quasi-time varying one, KWIK-based models perform better.
The MovieLens dataset \cite{Harper:2015:MDH:2866565.2827872} is used
to verify the models. The contribution of this paper is summarized
as follows:
\begin{enumerate}
\item For a PPP system, we find the optimum content caching probabilities
to maximize the ASP, when the popularity distribution is known.
\item We consider two scenarios \textendash{} time varying and quasi-time
varying. We present two main models \textendash{} PPM and GPM, and
three models for comparison \textendash{} IPM, RPM and ASP-PM. Based
on these models, online prediction, online learning, and KWIK framework
based approaches have been investigated. For GPM, the ASP regret is
analyzed for prediction models, while MSE regret is bounded for both
PPM and GPM. Simulations are performed and real dataset is utilized
to show the improved performance of these methods in time varying
scenarios.
\end{enumerate}
The organization of the paper is as follows: section \ref{sec:System-Model}
describes the system model. In section \ref{sec:Average-Success-Probability},
ASP has been maximized. For time-varying popularities, the next section
\ref{sec:Linear-Prediction-Models} explains the online prediction
methods, while the following section \ref{sec:Online-Learning-Models}
presents the online learning approaches. For quasi-time varying scenarios,
KWIK based learning methods are presented in section \ref{sec:Online-Learning-KWIK}.
Simulation results are provided in section \ref{sec:Simulation-Results}.
Section \ref{sec:Conclusion} concludes the paper. 

\section{System Model\label{sec:System-Model}}

Consider a cellular network where the positions of base stations (BSs)
are spatially distributed according to a two-dimensional (2D) homogeneous
Poisson point process (PPP) $\Phi_{BS}$ with density $\lambda_{bs}>0$.
Without loss of generality, from Slivanyak-Mecke theorem, for stationary
and homogenuity of PPP, we consider a typical user at the origin for
evaluating the performance. 

It is assumed that each user requests a content from the
content set $\mathcal{F}\coloneqq\left\{ f_{1},f_{2},\ldots,f_{N}\right\} $
of $N$ files. Each content is of same size and normalized to $1$.
Let $\mathbf{p}=[p_{1},\ldots,p_{N}]^{T}$ be the popularity distribution,
i.e., $\mathbf{p}^{T}\mathbf{1}=1$ and $\mathbf{p}\geq\mathbf{0}$.
In the given network, these popularities are computed as the normalized
number of file requests. In practice, the popularity profile follows
a Zipf probability mass function, i.e., the probability that a user
requests content $f_{j}$ is given as $p_{j}=\frac{j^{-\gamma}}{\sum_{i=1}^{N}i^{-\gamma}},\quad1\leq j\leq N,$
where $\gamma<1$ is Zipf exponent. Each BS has a cache memory of
size $L$. The cache memory at $i^{th}$ BS is denoted by $\mathcal{L}_{i}$,
which is a subset of $\mathcal{F}$, such that $\left|\mathcal{L}_{i}\right|\leq L$.
Assuming independent placement in caches, the probability that content
$f_{j}$ is stored a given BS as $q_{j}=\Pr\left[f_{j}\in\mathcal{L}_{i}\right],\quad1\leq j\leq N$
\cite{blaszczyszyn2015optimal}. The probability that a typical user
finds the desired content in a cache depends on the distribution of
the random set $\mathcal{L}_{i}$ \emph{only }through the one-set
coverage probabilities $q_{j}$. It defines the content placement
policy for the network as a whole. Hence, these probabilities satisfy
the cache constraint $\sum_{j=1}^{N}q_{j}\leq L.$ In \cite{blaszczyszyn2015optimal},
it has been shown that this condition is a necessary and sufficient
condition for the existence of a distribution on the random set $\mathcal{L}_{i}$
satisfying $\left|\mathcal{L}_{i}\right|\leq L$ almost surely.

\section{Average Success Probability (ASP) Maximization\label{sec:Average-Success-Probability}}

We consider user association based on both the channel state information
(CSI) and cached files in each helper. Specifically, when a user requests
$l^{th}$ file, it associates with the BS in the set $\Phi_{BS}(l)$
that has the strongest received power. Assuming equal power allocation
($P$), the received power at the user is $\left|h_{i}\right|^{2}Pr_{i}^{-\alpha}$,
where $\left|h_{i}\right|^{2}\sim\exp(1)$ is the channel gain, $r_{i}$
is the distance between $i^{th}$ BS and user, and $\alpha$ is the
path loss exponent. The downlink SINR at the typical user that request
$l^{th}$ file from $i^{th}$ BS is given as 
\begin{equation}
\Gamma_{il}=\frac{\left|h_{i}\right|^{2}Pr_{i}^{-\alpha}}{\underbrace{\sum_{j\in\Phi_{BS}\setminus\{i\}}\left|h_{j}\right|^{2}Pr_{j}^{-\alpha}}_{I_{i}=I_{il}+I_{il}^{c}}+\sigma^{2}},
\end{equation}
where $I_{il}=\sum_{j\in\Phi_{BS}(l)\setminus\{i\}}\frac{\left|h_{j}\right|^{2}P}{r_{j}^{\alpha}}$,
$I_{il}^{c}=I_{i}-I_{il}$, and $\sigma^{2}$ is the noise variance.
The term $I_{il}$ represents the interference from the BSs that cache
$l^{th}$ file, while $I_{il}^{c}$ is the interference from those
BSs which do not cache $l^{th}$ file. From the user's perspective,
we use success probability to reflect quality of service, which is
defined as the probability that the achievable rate of a typical user
exceeds the rate requirements $R_{0}$. The average success probability
can be written as 
\begin{equation}
P(\mathbf{p},\mathbf{q})=\sum_{l}p_{l}\Pr\left\{ W\log_{2}\left(1+\Gamma_{il}\right)\geq R_{0}\right\} ,
\end{equation}
where $W$ is the transmission bandwidth. The following result presents
the ASP expression. 
\begin{thm}
Average success probability of a typical user requesting $l^{th}$
file, which has popularity $p_{l}$ and caching probability $q_{l}$,
is given as $P(\mathbf{p},\mathbf{q})=\sum_{l=1}^{N}p_{l}g(q_{l}),$
where 
\begin{align}
g(q_{l}) & =q_{l}C\int_{0}^{\infty}\exp\left[-s_{0}r_{i}^{\alpha}\left(\frac{\sigma^{2}}{P}\right)\right]\nonumber \\
 & \quad\times\exp\left[-\left(q_{l}A+(1-q_{l})B+q_{l}C\right)r_{i}^{2}\right](dr_{i}^{2})\\
s_{0} & =2^{\frac{R_{0}}{W}}-1\\
A & =2\pi\lambda_{bs}s_{0}^{\frac{2}{\alpha}}\frac{1}{\alpha}\int_{\frac{1}{s_{0}}}^{\infty}\left(\frac{u^{\frac{2}{\alpha}-1}}{1+u}\right)du\\
B & =2\pi\lambda_{bs}s_{0}^{\frac{2}{\alpha}}\frac{1}{\alpha}\int_{0}^{\infty}\left(\frac{u^{\frac{2}{\alpha}-1}}{1+u}\right)du\\
C & =\pi\lambda_{bs}.
\end{align}
\end{thm}
\begin{IEEEproof}
Proof is given in Appendix\ref{subsec:ASP-Derivation}.
\end{IEEEproof}
Since heterogeneous networks are usually interference limited, it
is reasonable to neglect the noise i.e., $\sigma^{2}=0$. For this
case, the corollary below simplifies the ASP. 
\begin{cor}
For interference limited case, i.e., $\sigma^{2}=0$ or at high SNR,
average success probability is given as $P_{s}(\mathbf{p},\mathbf{q})=\sum_{l}p_{l}g_{0}(q_{l}),$
where 
\begin{align}
g_{0}(q_{l}) & =q_{l}C\int_{0}^{\infty}\exp\left[-\left(q_{l}A+(1-q_{l})B+q_{l}C\right)r_{i}^{2}\right](dr_{i}^{2})\nonumber \\
 & =\frac{q_{l}C}{q_{l}A+(1-q_{l})B+q_{l}C}.\label{eq:g0fun}
\end{align}
\end{cor}
Given the content popularity distribution, the next step is to place
the content in the caches. Since PPP based network is considered as
a whole, the optimal caching probabilities are computed in order to
maximize the ASP. The ASP maximization problem with respect to caching
can be expressed as 
\begin{align*}
 & \max_{\mathbf{q}}P_{s}(\mathbf{p},\mathbf{q})\\
 & \text{ }\qquad\text{subject to }\mathbf{q}^{T}\mathbf{1}\leq L.
\end{align*}

For simplicity, to be solvable in CVX tool \cite{Boyd2010}, the above
problem can be cast in semi-definite program (SDP) as 
\begin{align*}
 & \min_{\mathbf{q},\mathbf{t}}-\sum_{l}t_{l}p_{l}\\
 & \text{ }\qquad\text{subject to }\mathbf{q}^{T}\mathbf{1}\leq L\\
 & \left[\begin{array}{cc}
q_{l}(A-B+C)+B & 1\\
1 & \left(\frac{A-B+C}{BC}t-\frac{1}{B}\right)
\end{array}\right]\preceq\mathbf{0},\forall l
\end{align*}
where $\frac{q_{l}C}{q_{l}A+(1-q_{l})B+q_{l}C}\geq t_{l}$ has been
simplified using Schur's lemma \cite{Boyd2010}. Further, the expression
of the solution is presented in the following theorem. 
\begin{thm}
The solution of the maximization problem is given as 
\begin{align}
\mathbf{q}^{*} & =\arg\max_{\mathbf{q}^{T}\mathbf{1}\leq L}P_{s}(\mathbf{p},\mathbf{q})\\
 & =\begin{cases}
1, & i\in\mathcal{R}\\
\left(\frac{B}{A+C-B}\right)\left[\frac{\eta\bar{\mathbf{p}}}{\mathbf{1}^{T}\bar{\mathbf{p}}(\mathcal{P})}-1\right], & i\in\mathcal{P}\\
0, & i\in\mathcal{Z}
\end{cases}\label{eq:qOpt}
\end{align}
where $\bar{\mathbf{p}}=\sqrt{\mathbf{p}}$, $\mathcal{Z}=\left\{ i|q_{i}=0\right\} ,\mathcal{R}=\left\{ i|q_{i}=1\right\} $,
and $\mathcal{P}=\left[N\right]\setminus\left\{ \mathcal{Z}\cup\mathcal{R}\right\} =\left\{ i|0<q_{i}<1\right\} $.
The corresponding ASP is obtained as 
\[
P_{s}(\mathbf{p},\mathbf{q}^{*})=\left(\frac{C}{A+C-B}\right)\bar{\mathbf{p}}^{T}\bar{\mathbf{Z}}\bar{\mathbf{p}}
\]
where $\bar{\mathbf{Z}}=\mathcal{D}\left(\frac{A+C-B}{A+C}\mathbf{I}_{|\mathcal{R}|},\mathbf{Z}_{\mathcal{P}},\mathbf{0}_{|\mathcal{Z}|}\right)$,
$\mathbf{Z}_{\mathcal{P}}=\mathbf{I}-\frac{\mathbf{1}\mathbf{1}^{T}}{\eta}$
and $\eta=|\mathcal{P}|+(M-\left|\mathcal{R}\right|)\frac{A-B+C}{B}$. 
\end{thm}
\begin{IEEEproof}
Proof is given in Appendix\ref{subsec:Solution-to-maximization}.
\end{IEEEproof}
The above expression of $\mathbf{q}^{*}$ depends on the index sets
($\mathcal{Z}$ and $\mathcal{R}$), which are obtained in Algorithm
\ref{alg:Max-ASP-Algorithm}. This algorithm employs the inferences
obtained from Karush\textendash Kuhn\textendash Tucker (KKT) conditions
to get the index sets. In its first part, individual $l^{th}$ index
is checked for $q_{l}^{*}=1$, while the later part is for $q_{l}^{*}=0$.
Now, with the caching probabilities obtained for a PPP network, a
random caching strategy is utilized to place the content in individual
caches \cite{blaszczyszyn2015optimal}.
\begin{algorithm}
 \begin{algorithmic}[1] 
\Require{$\mathbf{p}$ in descending order}
\Ensure{$\mathcal{R}$, $\mathcal{Z}$ and $\mathcal{P}$}
\State{Initialize $\mathcal{R}=\emptyset$, $\mathcal{P}=\mathcal{N}\setminus\mathcal{R}$, $j=0$.}
\While{$\frac{B}{A+C}\bar{p}_{i}>\frac{\bar{\mathbf{p}}(\mathcal{P})^{T}\mathbf{1}}{\eta},\forall i\in\mathcal{R}$}
\State{$j=j+1$ and add $j$ in $\mathcal{R}$ and remove from $\mathcal{P}$}
\EndWhile
\State{Initialize $\mathcal{Z}=\emptyset$, $\mathcal{P}=\mathcal{N}\setminus\left\{ \mathcal{R}\cup\mathcal{Z}\right\} $, $j=0$.}
\While{$\frac{\bar{\mathbf{p}}(\mathcal{P})^{T}\mathbf{1}}{\eta}<\bar{p}_{i},\forall i\in\mathcal{Z}$}
\State{$j=j+1$ and add $j$ in $\mathcal{Z}$ and remove from $\mathcal{P}$}
\EndWhile
\end{algorithmic}

\caption{Getting Index sets for ASP maximization\label{alg:Max-ASP-Algorithm}}
\end{algorithm}

From the above approach, content placement solely depend on the content
popularity distribution. Therefore, it is important to predict this
popularity profile in order to cache contents in advance. The content
request prediction problem can be presented as follows. Let $n_{lt}$
denote the number of requesting content observed during the $t^{th}$
slot time for $l^{th}$ file. At time $t$, the task is to predict
requests $n_{l(t+i)}$ at time $t+i$ based on the past content request
sequence $\left\{ n_{lj}|j=1,\ldots,t\right\} $. In the following
section, based on the normalized content requests (popularity profile),
online prediction and online learning models are presented for estimating
the content probabilities considering time varying and quasi-time
varying scenarios.

\section{Online Prediction Models With Time Varying Popularities\label{sec:Linear-Prediction-Models}}

In this section, time varying popularity models are considered for
online predictions. Typically, in time series prediction, the data
of present time slot depends on the linear combination of the data
in previous time slots. Here, we refer to the scenario, where the
present content popularity is changing in each time slot and is independent
of profile in previous time slots. For this scenario, we fit linear
model and use online least squares to find time varying coefficients
of regression. In the following, in particular, two models are presented
that have shown better performance in terms of MSE and ASP. Moreover,
three models are briefly given for comparison. 

\subsection{Popularity Prediction Model (PPM)\label{subsec:PPM} }

Let $\mathbf{p}_{t}$ denote the content popularity vector in time
slot $t$ such that $\mathbf{p}_{t}^{T}\mathbf{1}=1$. In this model,
the present content popularity can be approximated as a linear combination
of past observed popularity vectors as 
\begin{align}
\mathbf{p}_{t} & \approx\sum_{k=1}^{d}c_{k}\mathbf{p}_{t-k},\label{eq:vecLP}
\end{align}
with $c_{k}\in\mathbb{R}$ are the coefficients of prediction satisfying
$\mathbf{p}_{t}^{T}\mathbf{1}=\sum_{k=1}^{d}c_{k}=1$. Note that the
constraint $c_{k}\in\mathbb{R}$ is necessary because if $c_{k}>0$,
\eqref{eq:vecLP} represents a convex combination of past popularities.
It means if content popularity is increasing/decreasing in the next
time slot, this model $\left(c_{k}>0\right)$ will not be able to
predict the popularity profile. Therefore, for known previous $\tau$
popularities $p_{l(t-\tau+1)},\,\ldots,\,p_{lt}$, the coefficients
can be estimated as 
\begin{align}
\mathbf{c} & =\arg\min_{c_{k}\in\mathbb{R}}\sum_{i=t-\tau+d+1}^{t}\frac{1}{2}\left\Vert \mathbf{p}_{i}-\sum_{k=1}^{d}c_{k}\mathbf{p}_{i-k}\right\Vert ^{2}\\
 & \text{subject to }\mathbf{1}^{T}\hat{\mathbf{p}}_{t+1}=\sum_{k=1}^{d}c_{k}\mathbf{1}^{T}\mathbf{p}_{t+1-k}=\sum_{k=1}^{d}c_{k}=1\label{eq:PPP_con_sum1}\\
 & \qquad\qquad0\leq\hat{\mathbf{p}}_{t+1}=\sum_{k=1}^{d}c_{k}\mathbf{p}_{t+1-k},\:
\end{align}
where $\hat{\mathbf{p}}_{t+1}$ is the predicted profile in $(t+1)^{th}$
time slot. In above formulation, the constraint in \eqref{eq:PPP_con_sum1}
ensures predicted popularities summable to one, while the next constraint
restricts the probabilities to be non-negative. Note that the constraint
$\mathbf{1}^{T}\hat{\mathbf{p}}_{t+1}\leq1$ is implicit as $\hat{\mathbf{p}}_{t+1}\geq\mathbf{0}$
and $\hat{\mathbf{p}}_{t+1}^{T}\mathbf{1}=1$ are sufficient to ensure
the upper bound on $\hat{\mathbf{p}}_{t+1}$. Further, the above optimization
can be simplified as 
\begin{align}
\hat{\mathbf{p}}_{t+1}= & \arg\min_{\mathbf{x}\geq\mathbf{0}}\frac{1}{2}\left\Vert \mathbf{y}_{t}-\mathbf{H}_{t}\mathbf{x}\right\Vert ^{2}\label{eq:LS_ppm}\\
 & \text{subject to }\mathbf{1}^{T}\mathbf{x}=1
\end{align}
where $\mathbf{y}_{t}=\left[\mathbf{p}_{t-\tau+d+1}^{T},\ldots,\mathbf{p}_{t}^{T}\right]^{T}$,
$\mathbf{x}=\sum_{k=1}^{d}c_{k}\mathbf{p}_{t+1-k}=\mathbf{P}_{t}\mathbf{c}$,
$\mathbf{H}'_{t}=\left[\mathbf{y}_{t-d},\ldots\mathbf{y}_{t-1}\right]$,
$\mathbf{H}_{t}=\mathbf{H}_{t}'\mathbf{P}_{t}^{-1}$. The above optimization
is a constrained non-negative least squares (NNLS) problem. NNLS without
additional constraint has been solved in \cite{lawson1995solving},
which provides an active set method, which is succeeded by \cite{971250f0a1b711ddb6ae000ea68e967b}.
\cite{971250f0a1b711ddb6ae000ea68e967b} modifies it to have a faster
version, called fast NNLS (FNNLS). Here, fast NNLS algorithm is modified
to obtain the solution for constrained NNLS in \eqref{eq:LS_ppm}.
Toward this, the Lagrangian and KKT conditions can be written as
\begin{align}
\mathcal{L} & =\frac{1}{2}\left\Vert \mathbf{y}_{t}-\mathbf{H}_{t}\mathbf{x}\right\Vert ^{2}+\lambda(\mathbf{1}^{T}\mathbf{x}-1)+\mathbf{v}^{T}\mathbf{x}.\\
\mathbf{0} & =\mathbf{H}_{t}^{T}\mathbf{H}_{t}\mathbf{x}-\mathbf{H}_{t}^{T}\mathbf{y}_{t}+\lambda\mathbf{1}+\mathbf{v},\\
0 & =\lambda(\mathbf{1}^{T}\mathbf{x}-1),\:\lambda\neq0\\
0 & =v_{i}x_{i},\:v_{i}\leq0,\,\forall i.
\end{align}
Solving the above equation gives 
\begin{align}
\mathbf{x}(\mathbf{H}_{t},\mathbf{v}) & =\left(\mathbf{H}_{t}^{T}\mathbf{H}_{t}\right)^{-1}\left(\mathbf{H}_{t}^{T}\mathbf{y}_{t}-\lambda(\mathbf{H}_{t},\mathbf{v})\mathbf{1}-\mathbf{v}\right),\label{eq:PPMsol1}\\
\lambda(\mathbf{H}_{t},\mathbf{v}) & =\frac{\mathbf{1}^{T}\left(\mathbf{H}_{t}^{T}\mathbf{H}_{t}\right)^{-1}\left(\mathbf{H}_{t}^{T}\mathbf{y}_{t}-\mathbf{v}\right)-1}{\mathbf{1}^{T}\left(\mathbf{H}_{t}^{T}\mathbf{H}_{t}\right)^{-1}\mathbf{1}}.\label{eq:PPMsol2}
\end{align}

\begin{algorithm}
\begin{algorithmic}[1] \Require{Initialize $\mathcal{P}=\emptyset$, $\mathcal{R}=\left\{ 1,\ldots,n\right\} $,  $\mathbf{x}=\mathbf{0}$,   $\mathbf{v}=\mathbf{H}_{t}^{T}(\mathbf{y}_{t}-\mathbf{H}_{t}\mathbf{x})$ and tolerance $\epsilon$} \Ensure{$\mathbf{x}\geq\mathbf{0}$ such that $\mathbf{1}^{T}\mathbf{x}=1$} 
\While{$\mathcal{R}\neq\emptyset$ and $\max_{i}v_{i}>\epsilon$} 
\State{set $j=\arg\max_{i}v_{i}$, add $j$ in $\mathcal{P}$ and remove from $\mathcal{R}$} \State{set $\mathbf{s}_{\mathcal{R}}=\mathbf{0}$ and $\mathbf{s}_{\mathcal{P}}=\mathbf{x}_{\mathcal{P}}(\mathbf{v})$,} 
\If{$\min\mathbf{s}_{\mathcal{P}}\leq0$} 
\State{$\alpha=-\min\left[\frac{\mathbf{x}_{\mathcal{P}}}{\mathbf{x}_{\mathcal{P}}-\mathbf{s}_{\mathcal{P}}}\right]$} 
\If{$\alpha=0$} 
\State{$\mathcal{I}=\{i|x_{i}\neq0\}$  and $\alpha=-\min\left[\frac{\mathbf{x}_{\mathcal{I}}}{\mathbf{x}_{\mathcal{I}}-\mathbf{s}_{\mathcal{I}}}\right]$} \EndIf \State{$\mathbf{x}=\mathbf{x}+\alpha(\mathbf{s}-\mathbf{x})$}
\State{update $\mathcal{P}=\{i|x_{i}>0\}$ and $\mathcal{R}=\left\{ i|x_{i}\leq0\right\} $ and set $\mathbf{s}_{\mathcal{R}}=\mathbf{0}$ and $\mathbf{s}_{\mathcal{P}}=\mathbf{x}_{\mathcal{P}}(\mathbf{v})$} \EndIf  
\State{$\mathbf{x}=\mathbf{s}$ and $\mathbf{v}=\mathbf{H}_{t}^{T}(\mathbf{y}_{t}-\mathbf{H}_{t}\mathbf{x})-\lambda_{\mathcal{P}}(\mathbf{v})\mathbf{1}$} \EndWhile \end{algorithmic} 

\caption{Modified fast NNLS algorithm\label{alg:Modified-NNLS-Algorithm}}
\end{algorithm}
In fast NNLS, the following equations are employed modified from
\eqref{eq:PPMsol1}-\eqref{eq:PPMsol2}
\[
\mathbf{x}_{\mathcal{P}}(\mathbf{v})=\left(\left(\mathbf{H}_{t}^{T}\mathbf{H}_{t}\right)^{\mathcal{P}}\right)^{-1}\left(\left(\mathbf{H}_{t}^{T}\mathbf{y}_{t}\right)^{\mathcal{P}}-\lambda_{\mathcal{P}}(\mathbf{H}_{t},\mathbf{v})\mathbf{1}-\mathbf{v}\right),
\]
\begin{align}
\lambda_{\mathcal{P}}(\mathbf{v}) & =\frac{\mathbf{1}^{T}\left(\left(\mathbf{H}_{t}^{T}\mathbf{H}_{t}\right)^{\mathcal{P}}\right)^{-1}\left(\left(\mathbf{H}_{t}^{T}\mathbf{y}_{t}\right)^{\mathcal{P}}-\mathbf{v}\right)-1}{\mathbf{1}^{T}\left(\left(\mathbf{H}_{t}^{T}\mathbf{H}_{t}\right)^{\mathcal{P}}\right)^{-1}\mathbf{1}}.
\end{align}
The feature of above equations is that they compute the matrix multiplications
($\mathbf{H}_{t}^{T}\mathbf{H}_{t}$ and $\mathbf{H}_{t}^{T}\mathbf{y}_{t}$)
once. Algorithm \ref{alg:Modified-NNLS-Algorithm} presents the required
procedure to obtain the solution, where the number of ``while''
iterations is equal to the number of non-zero entries of $\mathbf{x}$. 

\subsection{Grassmannian Prediction Model (GPM)}

From ASP optimization in \eqref{eq:qOpt}, it can be noted that positive
square root of caching probabilities maximizes ASP. The positive square
root of popularity vector represents a line in Grassmannian manifold
$\mathcal{G}_{N,1}$ \cite{el2012grassmannian,zhang2012robust}. This
model is presented as 

\begin{align}
\bar{\mathbf{p}}_{t} & \approx\sum_{k=1}^{d}z_{k}\bar{\mathbf{p}}_{t-k}
\end{align}
where $\bar{\mathbf{p}}_{t}=\sqrt{\mathbf{p}_{t}}$. Similar to
PPM, the coefficients are given using least squares as 
\begin{align}
\mathbf{z} & =\arg\min_{z_{k}\in\mathbb{R}}\sum_{i=t-\tau+d+1}^{t}\frac{1}{2}\left\Vert \bar{\mathbf{p}}_{i}-\sum_{k=1}^{d}z_{k}\bar{\mathbf{p}}_{i-k}\right\Vert ^{2}\\
 & \text{subject to }\left\Vert \hat{\bar{\mathbf{p}}}_{t+1}\right\Vert =\left\Vert \sum_{k=1}^{d}z_{k}\bar{\mathbf{p}}_{n+1-k}\right\Vert \leq1\\
 & \qquad\qquad0\leq\hat{\bar{\mathbf{p}}}_{t+1}=\sum_{k=1}^{d}z_{k}\bar{\mathbf{p}}_{n+1-k}\leq1,\:
\end{align}
By change of variables, the above problem can be recast as 
\begin{align}
\hat{\bar{\mathbf{p}}}_{t+1}= & \arg\min_{\mathbf{x}\geq\mathbf{0}}\frac{1}{2}\left\Vert \mathbf{y}_{t}-\mathbf{H}_{t}\mathbf{x}\right\Vert ^{2}\\
 & \text{subject to }\|\mathbf{x}\|_{2}^{2}\leq1
\end{align}
where $\mathbf{y}_{t}=\left[\bar{\mathbf{p}}_{t-\tau+d+1}^{T},\ldots,\bar{\mathbf{p}}_{t}^{T}\right]^{T}$,
$\mathbf{x}=\sum_{k=1}^{d}c_{k}\bar{\mathbf{p}}_{t+1-k}=\bar{\mathbf{P}}_{t}\mathbf{c}$,
$\mathbf{H}'_{t}=\left[\mathbf{y}_{t-d},\ldots\mathbf{y}_{t-1}\right]$,
$\mathbf{H}_{t}=\mathbf{H}_{t}'\bar{\mathbf{P}}_{t}^{-1}$. The above
problem is a regularized NNLS, whose Lagrangian and KKT conditions
can be written as 
\begin{align}
\mathcal{L} & =\frac{1}{2}\left\Vert \mathbf{y}_{t}-\mathbf{H}_{t}\mathbf{x}\right\Vert ^{2}+\lambda(\|\mathbf{x}\|_{2}^{2}-1)+\mathbf{v}^{T}\mathbf{x}\\
 & \mathbf{0}=\mathbf{H}_{t}^{T}\mathbf{H}_{t}\mathbf{x}-\mathbf{H}_{t}^{T}\mathbf{y}_{t}+\lambda\mathbf{x}+\mathbf{v},\\
 & 0=\lambda(\mathbf{x}^{T}\mathbf{x}-1),\lambda\geq0\\
 & 0=v_{i}x_{i},v_{i}\leq0,\forall i.
\end{align}
yielding $\mathbf{x}=\left(\mathbf{H}_{t}^{T}\mathbf{H}_{t}+\lambda\mathbf{I}\right)^{-1}\left(\mathbf{H}_{t}^{T}\mathbf{y}_{t}-\mathbf{v}\right)$.
For known $\mathbf{v}$, bisection can be employed to get $\lambda$
satisfying norm constraint. And Algorithm \ref{alg:Modified-NNLS-Algorithm}
is utilized to get $\mathbf{v}$. Alternatively, CVX tool can be used
for simplicity. 

Since the vector $\bar{\mathbf{p}}$ is related to ASP maximization,
here the optimum content placement is derived in terms of past placements.
At $t$ time slot, after updating $\mathcal{R}_{t}$, $\mathcal{P}_{t}$
and $\mathcal{Z}_{t}$ using Algorithm \ref{alg:Max-ASP-Algorithm},
content placement probabilities are updated for $i\in\mathcal{P}_{t}$
as 

\begin{align}
 & \hat{\mathbf{q}}_{t}(\mathcal{P}_{t})\left(\frac{B}{A+C-B}\right)^{-1}=\left[\frac{\eta_{t}\hat{\bar{\mathbf{p}}}_{i,t}(\mathcal{P}_{t})}{\mathbf{1}^{T}\hat{\bar{\mathbf{p}}}_{t}(\mathcal{P}_{t})}-\mathbf{1}\right]\\
 & =\left[\frac{\eta_{t}\left(z_{t-1}\bar{\mathbf{p}}_{t-1}(\mathcal{P}_{t})+\kappa_{t-1}\hat{\bar{\mathbf{p}}}_{t-1}(\mathcal{P}_{t})\right)}{\mathbf{1}^{T}\hat{\bar{\mathbf{p}}}_{t}(\mathcal{\left(\frac{B}{A+C-B}\right)P}_{t})}-\mathbf{1}\right]\\
 & =\frac{\eta_{t}\frac{z_{t-1}}{\kappa_{t}}\mathbf{1}^{T}\bar{\mathbf{p}}_{t-1}(\mathcal{P}_{t})}{\mathbf{1}^{T}\hat{\bar{\mathbf{p}}}_{t}(\mathcal{P}_{t})\eta_{t-1}}\left[\frac{\eta_{t-1}\bar{\mathbf{p}}_{t-1}(\mathcal{P}_{t})}{\mathbf{1}^{T}\bar{\mathbf{p}}_{t-1}(\mathcal{P}_{t})}-\mathbf{1}\right]+\frac{\eta_{t}\frac{\kappa_{t-1}}{\kappa_{t}}\mathbf{1}^{T}\hat{\bar{\mathbf{p}}}_{t-1}(\mathcal{P}_{t})}{\mathbf{1}^{T}\hat{\bar{\mathbf{p}}}_{t}(\mathcal{P}_{t})\eta_{t-1}}\left[\frac{\eta_{t-1}\hat{\bar{\mathbf{p}}}_{t-1}(\mathcal{P}_{t})}{\mathbf{1}^{T}\hat{\bar{\mathbf{p}}}_{t-1}(\mathcal{P}_{t})}-\mathbf{1}\right]\nonumber \\
 & -\left[1-\frac{\eta_{t}\frac{z_{t-1}}{\kappa_{t}}\mathbf{1}^{T}\bar{\mathbf{p}}_{t-1}(\mathcal{P}_{t})}{\mathbf{1}^{T}\hat{\bar{\mathbf{p}}}_{t}(\mathcal{P}_{t})\eta_{t-1}}-\frac{\eta_{t}\frac{\kappa_{t-1}}{\kappa_{t}}\mathbf{1}^{T}\hat{\bar{\mathbf{p}}}_{t-1}(\mathcal{P}_{t})}{\mathbf{1}^{T}\hat{\bar{\mathbf{p}}}_{t}(\mathcal{P}_{t})\eta_{t-1}}\right]\mathbf{1}\\
 & \hat{\mathbf{q}}_{t}(\mathcal{P}_{t})=\omega_{1,t-1}\mathbf{q}_{t-1}(\mathcal{P}_{t})+\omega_{2,t-1}\hat{\mathbf{q}}_{t-1}(\mathcal{P}_{t})+\left(\frac{B}{A+C-B}\right)(1-\omega_{1,t-1}-\omega_{2,t-1})\mathbf{1}\nonumber 
\end{align}
where $\omega_{1,t-1}=\frac{\eta_{t}\frac{z_{t-1}}{\kappa_{t}}\mathbf{1}^{T}\bar{\mathbf{p}}_{t-1}(\mathcal{P}_{t})}{\mathbf{1}^{T}\hat{\bar{\mathbf{p}}}_{t}(\mathcal{P}_{t})\eta_{t-1}}$,
and $\omega_{2,t-1}=\frac{\eta_{t}\frac{\kappa_{t-1}}{\kappa_{t}}\mathbf{1}^{T}\hat{\bar{\mathbf{p}}}_{t-1}(\mathcal{P}_{t})}{\mathbf{1}^{T}\hat{\bar{\mathbf{p}}}_{t}(\mathcal{P}_{t})\eta_{t-1}}$.

ASP difference with respect to optimal one (when popularity is known
perfectly) can written as $\Delta_{t}=P_{s}(\mathbf{p}_{t},\mathbf{q}_{t})-P_{s}(\mathbf{p}_{t},\hat{\mathbf{q}}_{t})=\left(\frac{C}{A+C-B}\right)\bar{\mathbf{p}}_{t}^{T}\left[\bar{\mathbf{Z}}_{t}-\hat{\mathbf{Z}}_{t}\right]\bar{\mathbf{p}}_{t}$,
\begin{figure*}
\begin{align}
 & \Delta_{t}=\left(\frac{C}{A+C-B}\right)\bar{\mathbf{p}}_{t}^{T}\mathcal{D}\left(\frac{-B}{A+C}\mathbf{I}_{|\mathcal{R}_{t}|},-\mathbf{1}\mathbf{1}^{T}+\|\bar{\mathbf{p}}_{t}(\mathcal{P}_{t})\|_{1}\mathcal{D}^{-1}\left(\bar{\mathbf{p}}_{t}(\mathcal{P}_{t})\right),\mathbf{0}_{|\mathcal{Z}_{t}|}\right)\bar{\mathbf{p}}_{t}\\
 & \leq\left(\frac{C}{A+C-B}\right)\frac{1}{\eta}\lambda_{\max}\left[-\mathbf{1}\mathbf{1}^{T}+\|\bar{\mathbf{p}}_{t}(\mathcal{P}_{t})\|_{1}\mathcal{D}^{-1}\left(\bar{\mathbf{p}}_{t}(\mathcal{P}_{t})\right)\right]-\frac{B}{A+C}\sum_{i\in\mathcal{R}_{t}}p_{i,t}\\
 & =\left(\frac{C}{A+C-B}\right)\frac{1}{\eta}\lambda_{\max}\left[\frac{\mathbf{1}}{\sqrt{\left|\mathcal{P}_{t}\right|}}\left\{ -\left|\mathcal{P}_{t}\right|+\|\bar{\mathbf{p}}_{t}(\mathcal{P}_{t})\|_{1}\mathcal{D}^{-1}\left(\bar{\mathbf{p}}_{t}(\mathcal{P}_{t})\right)\right\} \frac{\mathbf{1}^{T}}{\sqrt{\left|\mathcal{P}_{t}\right|}}\right]-\frac{B}{A+C}\sum_{i\in\mathcal{R}_{t}}p_{i,t}\nonumber \\
 & =\left(\frac{C}{A+C-B}\right)\frac{1}{\eta}\left\{ -\left|\mathcal{P}_{t}\right|+\frac{\sum_{i\in\mathcal{P}_{t}}\bar{p}_{i,t}}{\min_{i\in\mathcal{P}_{t}}\bar{p}_{i,t}}\right\} -\frac{B}{A+C}\sum_{i\in\mathcal{R}_{t}}p_{i,t}\label{eq:DeltaPs-1}
\end{align}
\end{figure*}
where $\bar{\mathbf{Z}}_{t}=\mathcal{D}\left(\frac{A+C-B}{A+C}\mathbf{I}_{|\mathcal{R}_{t}|},\mathbf{Z}_{\mathcal{P}_{t}},\mathbf{0}_{|\mathcal{Z}_{t}|}\right)$
and $\hat{\mathbf{Z}}_{t}=\mathcal{D}\left(\mathbf{I}_{|\hat{\mathcal{R}}_{t}|},\mathbf{I}-\|\hat{\bar{\mathbf{p}}}_{t}(\hat{\mathcal{P}}_{t})\|_{1}\mathcal{D}^{-1}\left(\hat{\bar{\mathbf{p}}}_{t}(\hat{\mathcal{P}}_{t})\right),\mathbf{0}_{|\hat{\mathcal{Z}}_{t}|}\right)$.
Assuming that prediction model results in at least accurate content
placement indices sets i.e., $\mathcal{R}_{t}=\hat{\mathcal{R}}_{t}$
and $\mathcal{P}_{t}=\hat{\mathcal{P}}_{t}$. Therefore, one can
simply above and get \eqref{eq:DeltaPs-1} given at the top of page,
where we used $\lambda_{\max}\left[\mathbf{v}\mathbf{v}^{T}+\mathbf{D}\right]=\lambda_{\max}\left[\mathbf{v}\left(\mathbf{I}+\mathbf{D}\right)\mathbf{v}^{T}\right]=\mathbf{I}+\max(\mathbf{D})$
for a diagonal matrix $\mathbf{D}$ and $\|\mathbf{v}\|=1$. 

When the popularities are not known ($p_{\ensuremath{l,t}}=\hat{p}_{\ensuremath{l,t}}+\delta_{\ensuremath{l,t}}$),
the ASP regret is obtained as
\begin{align}
\hat{\Delta}_{t} & =P_{s}(\mathbf{p}_{t},\mathbf{q}_{t})-P_{s}(\hat{\mathbf{p}}_{t},\hat{\mathbf{q}}_{t})\\
 & =\sum_{l}\left(p_{l,t}g_{0}(q_{l,t})-\hat{p}_{l,t}g_{0}(\hat{q}_{l,t})\right)\\
 & =\sum_{l}\left(p_{l,t}g_{0}(q_{l,t})-p_{l,t}g_{0}(\hat{q}_{l,t})-\delta{}_{l,t}g_{0}(\hat{q}_{l,t})\right)\\
 & =\Delta_{t}-P_{s}(\mathbf{\delta}_{t},\hat{\mathbf{q}}_{t})\\
 & =\Delta_{t}-\left(\frac{C}{A+C-B}\right)\left(\sum_{l\in\mathcal{R}_{t}}\delta_{l,t}+\sum_{l\in\mathcal{P}_{t}}\delta_{l,t}\left(1-\frac{\sum_{j\in\mathcal{P}_{t}}\hat{\bar{p}}_{j,t}}{\hat{\bar{p}}_{l,t}}\right)\right)
\end{align}

The above bound shows that the prediction achieves constant gap for
average success probability. It also shows that as the number of files
increases the bound becomes tighter. By adjusting the physical layer
parameters, one can achieve arbitrarily close bound.

\subsection{Models For Comparison }

In this section, three different models are presented for comparison.
These models with some minor variations have been employed in the
literature \cite{yin2018prediction,liu2018content,nakayama2015caching,zhang2018ppc}. 

\subsubsection{Product AR (log-request) Model}

This is the simplest model. In this model, the logarithm of requests
for each content is modeled as AR process. This model is expressed
as 
\begin{align}
\bar{n}_{lt} & =\log\frac{n_{lt}}{n_{\max}}=\sum_{k=1}^{d}c_{lk}\bar{n}_{l(t-k)}+\epsilon_{lt},
\end{align}
where $c_{lk}\in\mathbb{R}$, $\epsilon_{lt}$ is Gaussian error with
zero mean and variance $\sigma_{e}^{2}$, and $n_{\max}$ is known
value denoting the maximum number of requests in a given time slot.
The predicted number of requests can be obtained as the rounded values
of $\hat{n}_{lt}$, i.e., 
\begin{align}
\hat{n}_{l(t+1)} & =\left\lfloor \exp\left(\bar{n}_{lt}\log n_{\max}\right)\right\rfloor =\left\lfloor \prod_{k=1}^{d}n_{l(t+1-k)}^{c_{lk}}\right\rfloor 
\end{align}
Employing least squares approach for known $\tau$ requests $n_{l(t-\tau+1)},\,\ldots,\,n_{lt}$,
we have 
\begin{align*}
c_{lk} & =\arg\min_{c_{lk}}\sum_{i=t-\tau+d+1}^{t}\left(\bar{n}_{li}-\sum_{k=1}^{d}c_{lk}\bar{n}_{l(i-k)}\right)^{2}
\end{align*}
where $\tau\geq d$. The above equation is an unconstrained least
squares minimization, whose solution can be easily obtained. 

\subsubsection{Information Prediction Model (IPM)}

It can be noted that popularities are non-negative and summed to one.
In information theory, the value $-\log_{2}p_{i}$ represents the
``information'' for $i^{th}$ content. Therefore, to model prediction
of content popularities, this logarithm based model can be regarded
as \emph{information prediction model} (IPM). Let $\tilde{\mathbf{p}}\coloneqq-\log\mathbf{p}$,
then, this model can be expressed as 
\begin{align}
\tilde{\mathbf{p}}_{t} & =\sum_{k=1}^{d}c_{k}\tilde{\mathbf{p}}_{t-k}+\mathbf{e}_{t}
\end{align}
where $c_{k}\in\mathbb{R}$ and $\mathbf{e}_{t}$ is a non-negative
random number. Unconstrained least squares problem can be written
for known $\tau$ profiles $\mathbf{p}_{(t-\tau+1)},\,\ldots,\,\mathbf{p}_{t}$
as 
\begin{align}
\mathbf{c} & =\arg\min_{c_{k}}\sum_{i=t-\tau+d+1}^{t}\left\Vert \tilde{\mathbf{p}}_{i}-\sum_{k=1}^{d}c_{k}\tilde{\mathbf{p}}_{i-k}\right\Vert ^{2}.
\end{align}
Similar to the log-request model, it is an unconstrained LS problem
and the solution can be obtained.

\subsubsection{ASP Based Prediction Model (ASP-PM)}

This model is also important, since the objective of prediction is
to maximize ASP. In this model, the popularity profile is predicted
from the ASP in the previous time slots. By multiplying $g_{0}(q_{l})$
in \eqref{eq:vecLP} and summing over files results into ASP based
prediction model as 
\begin{align}
P_{s,t} & \approx\sum_{k=1}^{d}c_{k}P_{s,t-k}
\end{align}
with the similar conditions that $\sum_{k}c_{k}=1$ and $c_{k}\in\mathbb{R}$.
Similar to PPM, the optimization problem to obtain the coefficients
can be formulated in constrained NNLS as 
\begin{align}
\mathbf{c} & =\arg\min_{c_{k}\in\mathbb{R}}\sum_{i=t-\tau+d+1}^{t}\left\Vert P_{s,i}-\sum_{k=1}^{d}c_{k}P_{s,i-k}\right\Vert ^{2}\\
 & \text{subject to }\sum_{k=1}^{d}c_{k}=1,\,\sum_{k=1}^{d}c_{k}\mathbf{p}_{n-k}\geq\mathbf{0},
\end{align}
whose solution can be obtained from Algorithm \ref{alg:Modified-NNLS-Algorithm}.

\section{Online Learning Models for Time Varying Popularities\label{sec:Online-Learning-Models}}

In this section, for time varying popularity, we present online learning
methods to minimize the weighted cumulative loss. 

\subsection{Popularity Prediction Model (PPM) }

In this model, the goal of online learning is to minimize the weighted
sum of $l_{2}$ loss function 
\begin{equation}
\hat{\mathbf{p}}_{t}=\arg\min_{\mathbf{p}}\frac{1}{2}\sum_{i=1}^{t-1}c_{i}\|\mathbf{p}-\mathbf{p}_{i}\|^{2},
\end{equation}
where $\sum_{i=1}^{t-1}c_{i}=1$. A trivial selection for these coefficients
is $c_{i}=\frac{1}{t-1}$, $1\leq i\leq t-1$. The above optimization
leads to the model
\begin{align}
\hat{\mathbf{p}}_{t} & =\sum_{i=0}^{t-1}c_{i}\mathbf{p}_{i}\\
 & =c_{t-1}\mathbf{p}_{t-1}+\left(\sum_{i=0}^{t-2}c_{i}\right)\sum_{i=0}^{t-2}\frac{c_{i}}{\sum_{i=0}^{t-2}c_{i}}\mathbf{p}_{i}\\
 & =c_{t-1}\mathbf{p}_{t-1}+\left(1-c_{t-1}\right)\hat{\mathbf{p}}_{t-1}.\label{eq:pt_iter}
\end{align}
where for $c_{i}$, $\frac{c_{i}}{\sum_{i=0}^{t-2}c_{i}}=\frac{(t-1)^{-1}}{\sum_{i=0}^{t-2}(t-1)^{-1}}=(t-2)^{-1}$.
The above equation chooses a balance between the recent prediction
and recent observation. Equation \eqref{eq:pt_iter} can be written
as 
\begin{align}
\hat{\mathbf{p}}_{t}-\mathbf{p}_{t-1} & =\left(1-c_{t-1}\right)\left(\hat{\mathbf{p}}_{t-1}-\mathbf{p}_{t-1}\right).
\end{align}
Therefore, the difference with respect to optimal $\mathbf{p}^{*}$
can be given as 

\begin{align}
\frac{1}{2}\|\hat{\mathbf{p}}_{t-1}-\mathbf{p}_{t-1}\|^{2}-\frac{1}{2}\|\mathbf{p}^{*}-\mathbf{p}_{t-1}\|^{2} & \leq\frac{1}{2}\|\hat{\mathbf{p}}_{t-1}-\mathbf{p}_{t-1}\|^{2}-\frac{1}{2}\|\hat{\mathbf{p}}_{t}-\mathbf{p}_{t-1}\|^{2}\nonumber \\
 & =\frac{1}{2}\|\hat{\mathbf{p}}_{t-1}-\mathbf{p}_{t-1}\|^{2}\left[1-\left(1-c_{t-1}\right)^{2}\right]\\
 & \leq c_{t-1}\|\hat{\mathbf{p}}_{t-1}-\mathbf{p}_{t-1}\|^{2}<2c_{t-1}\nonumber 
\end{align}
where first inequality comes from \cite{shalev2012online}; and the
last inequality arises from triangle inequality. Regret is given as
\begin{align*}
 & \frac{1}{2}\sum_{i=1}^{T}c_{i}\|\hat{\mathbf{p}}_{i}-\mathbf{p}_{i}\|^{2}-\frac{1}{2}\sum_{i=1}^{T}c_{i}\|\mathbf{p}^{*}-\mathbf{p}_{i}\|^{2}<2\sum_{i=1}^{T}c_{i}^{2}\leq2\left(2-T^{-1}\right)
\end{align*}
where $\sum_{i=1}^{T}i^{-2}\leq\int_{1}^{T}t^{-2}dt=\int_{T^{-1}}^{1}du=2-T^{-1}$.
The procedure for online learning is given in Algorithm \ref{alg:Online-learning-PPM}.
In each $t^{th}$ step, prediction is obtained and the true popularity
is observed. 
\begin{algorithm}
\begin{algorithmic}[1]

\For{for $t=1,2,\ldots$ }

\State{set $c_{t}=t^{-1}$ and predict $\hat{\mathbf{p}}_{t+1}=c_{t}\mathbf{p}_{t}+\left(1-c_{t}\right)\hat{\mathbf{p}}_{t}$}

\State{observe $\mathbf{p}_{t+1}$ }

\EndFor

\end{algorithmic} 

\caption{Online learning algorithm for PPM. \label{alg:Online-learning-PPM}}
\end{algorithm}

\subsection{Grassmannian Prediction Model (GPM) }

Similar to PPM based online learning, the goal here is to minimize
the following loss
\begin{align}
\hat{\bar{\mathbf{p}}}_{t} & =\arg\min_{\|\bar{\mathbf{p}}\|\leq1}\frac{1}{2}\sum_{i=1}^{t-1}z_{i}\|\bar{\mathbf{p}}-\bar{\mathbf{p}}_{i}\|^{2},\\
 & =\sum_{i=0}^{t-1}\frac{z_{i}}{\left\Vert \sum_{i=0}^{t-1}z_{i}\bar{\mathbf{p}}_{i}\right\Vert }\bar{\mathbf{p}}_{i}
\end{align}
where $z_{i}\geq0$ and it leads to the model as 
\begin{align}
\hat{\bar{\mathbf{p}}}_{t} & =\sum_{i=0}^{t-1}\frac{z_{i}}{\kappa_{t}}\bar{\mathbf{p}}_{i}\\
 & =\frac{z_{t-1}}{\kappa_{t}}\bar{\mathbf{p}}_{t-1}+\frac{\kappa_{t-1}}{\kappa_{t}}\sum_{i=0}^{t-2}\frac{z_{i}}{\kappa_{t-1}}\bar{\mathbf{p}}_{i}\\
 & =\frac{z_{t-1}}{\kappa_{t}}\bar{\mathbf{p}}_{t-1}+\frac{\kappa_{t-1}}{\kappa_{t}}\hat{\bar{\mathbf{p}}}_{t-1},\label{eq:pt_iter-GPM}
\end{align}
with  $\kappa_{t}=\left\Vert \sum_{i=1}^{t-1}z_{i}\bar{\mathbf{p}}_{i}\right\Vert $.
Note that there is no upper bound constraint on $z_{i}$. Therefore,
we trivially choose $z_{t}=1-t^{-1}\geq0$, which provides 
\begin{align}
\kappa_{t}^{2}=\sum_{i=1}^{t-1}\sum_{j=1}^{t-1}\left(1-i^{-1}\right)\left(1-j^{-1}\right)\bar{\mathbf{p}}_{i}^{T}\bar{\mathbf{p}}_{j} & \leq\sum_{i=1}^{t-1}\sum_{j=1}^{t-1}\left(1-i^{-1}\right)\left(1-j^{-1}\right)=\left(\sum_{i=1}^{t-1}z_{i}\right)^{2}\nonumber \\
 & \leq\left[(t-1)-1-\log(t-1)\right]^{2}
\end{align}
where Cauchy Schwarz's inequality has been used. Equation \ref{eq:pt_iter-GPM}
can be given as 
\begin{align}
\hat{\bar{\mathbf{p}}}_{t}-\bar{\mathbf{p}}_{t-1} & =\left(\frac{z_{t-1}}{\kappa_{t}}-1\right)\bar{\mathbf{p}}_{t-1}+\frac{\kappa_{t-1}}{\kappa_{t}}\hat{\bar{\mathbf{p}}}_{t-1}
\end{align}
 Using above, the difference with respect to optimal $\mathbf{p}^{*}$
can be bounded as 
\begin{align}
\frac{1}{2}\|\hat{\bar{\mathbf{p}}}_{t-1}-\bar{\mathbf{p}}_{t-1}\|^{2}-\frac{1}{2}\|\bar{\mathbf{p}}^{*}-\bar{\mathbf{p}}_{t-1}\|^{2} & \leq\frac{1}{2}\|\hat{\bar{\mathbf{p}}}_{t-1}-\bar{\mathbf{p}}_{t-1}\|^{2}-\frac{1}{2}\|\hat{\bar{\mathbf{p}}}_{t}-\bar{\mathbf{p}}_{t-1}\|^{2}\\
 & =\frac{1}{2}\|\hat{\bar{\mathbf{p}}}_{t-1}-\bar{\mathbf{p}}_{t-1}\|^{2}-\frac{\left\Vert \left(z_{t-1}-\kappa_{t}\right)\bar{\mathbf{p}}_{t-1}+\kappa_{t-1}\hat{\bar{\mathbf{p}}}_{t-1}\right\Vert ^{2}}{2\kappa_{t}^{2}}\nonumber \\
 & \leq1-\frac{1}{2\kappa_{t}^{2}}\left|\kappa_{t-1}^{2}\|\hat{\bar{\mathbf{p}}}_{t-1}\|^{2}-\left(-z_{t-1}+\kappa_{t}\right)^{2}\|\bar{\mathbf{p}}_{t-1}\|^{2}\right|\\
 & =1-\frac{\left|\kappa_{t-1}^{2}-\left(z_{t-1}-\kappa_{t}\right)^{2}\right|}{2\kappa_{t}^{2}}\leq1-\frac{z_{t-1}}{\kappa_{t}}
\end{align}
where triangle inequality has been used, and approximated with $z_{t}^{2}\ll1$
and $\kappa_{t-1}\approx\kappa_{t}$ for large $t$. The regret can
be obtained as 
\begin{align}
\frac{1}{2}\sum_{i=1}^{T}z_{i}\|\hat{\bar{\mathbf{p}}}_{i}-\bar{\mathbf{p}}_{i}\|^{2}-\frac{1}{2}\sum_{i=1}^{T}z_{i}\|\bar{\mathbf{p}}^{*}-\bar{\mathbf{p}}_{i}\|^{2} & \leq\sum_{i=1}^{T}z_{i}-\frac{z_{i}^{2}}{\kappa_{i+1}}\\
 & <\sum_{i=1}^{T}z_{i}=\sum_{i=1}^{T}\left(1-i^{-1}\right)\leq T-\log T-1
\end{align}
The corresponding online learning procedure is presented in Algorithm
\ref{alg:Online-learning-GPM}.
\begin{algorithm}
\begin{algorithmic}[1]

\For{$t=1,2,\ldots$ }

\State{set $z_{t}=1-t^{-1}$ and predict $\hat{\bar{\mathbf{p}}}_{t+1}=z_{t}\bar{\mathbf{p}}_{t}+\kappa_{t}\hat{\bar{\mathbf{p}}}_{t},$}

\State{normalize $\hat{\bar{\mathbf{p}}}_{t+1}\leftarrow\frac{\hat{\bar{\mathbf{p}}}_{t+1}}{\|\hat{\bar{\mathbf{p}}}_{t+1}\|}$
and set $\kappa_{t+1}=\|\hat{\bar{\mathbf{p}}}_{t+1}\|$}

\State{observe $\bar{\mathbf{p}}_{t+1}$ }

\EndFor

\end{algorithmic} 

\caption{Online learning algorithm for GPM. \label{alg:Online-learning-GPM}}
\end{algorithm}

\subsection{Product AR (log-request) Model}

For this model, the following loss function is minimized
\[
\hat{n}_{lt}=\arg\min_{n_{l}\geq0}\sum_{i=1}^{t-1}\sum_{l=1}^{N}c_{i}\|\bar{n}_{li}-\log\frac{n_{l}}{n_{\max}}\|^{2}
\]
where $\bar{n}_{lt}=\log\frac{n_{lt}}{n_{\max}}$, $\sum_{i=1}^{t-1}c_{i}=1$,
and the predicted number of requests can be obtained as the rounded
values of $\hat{n}_{lt}$, i.e., 
\begin{align}
\hat{n}_{lt}=\left\lfloor n_{\max}\exp\left(\sum_{i=1}^{t-1}c_{i}\bar{n}_{li}\right)\right\rfloor  & =\left\lfloor n_{\max}\exp\left(c_{(t-1)}\hat{\bar{n}}_{l(t-1)}+(1-c_{(t-1)})\bar{n}_{lt}\right)\right\rfloor \\
 & =\left\lfloor \hat{n}_{l(t-1)}^{c_{(t-1)}}\times n_{lt}^{(1-c_{(t-1)})}\right\rfloor .
\end{align}
The last equation is the required term for online learning with some
optimal $c_{t}$. However, for simplicity, we set $c_{t}=t^{-1}$,
and analyze the regret similar to PPM as 
\begin{align}
\frac{1}{2}\sum_{i=1}^{T}\sum_{l=1}^{N}c_{i}\left[\|\bar{n}_{li}-\hat{\bar{n}}_{li}\|^{2}-\|\bar{n}_{li}-\log\frac{n_{l}^{*}}{n_{\max}}\|^{2}\right] & \leq\frac{1}{2}\sum_{i=1}^{T}\sum_{l=1}^{N}c_{i}\left[\|\bar{n}_{li}-\hat{\bar{n}}_{li}\|^{2}-\|\bar{n}_{li}-\hat{\bar{n}}_{l(i-1)}\|^{2}\right]\nonumber \\
 & =\frac{1}{2}\sum_{i=1}^{T}\sum_{l=1}^{N}c_{i}\|\bar{n}_{lt}-\hat{\bar{n}}_{lt}\|^{2}\left[1-\left(1-c_{i}\right)^{2}\right]\nonumber \\
 & \leq\sum_{i=1}^{T}\sum_{l=1}^{N}c_{i}^{2}(\bar{n}_{lt}^{2}+\hat{\bar{n}}_{lt}^{2})\\
 & \leq\sum_{i=1}^{T}2Nc_{i}^{2}\log n_{\max}\leq2N\log n_{\max}\left(2-T^{-1}\right)\nonumber 
\end{align}
where $\left|\bar{n}_{lt}\right|=\left|\log\frac{n_{lt}}{n_{\max}}\right|\leq\log n_{\max}$.

\subsubsection{Information Prediction Model (IPM)}

Similar to the previous case, the learning objective can be expressed
as 
\begin{equation}
\tilde{\mathbf{p}}_{t}=\arg\min_{\mathbf{p}}\frac{1}{2}\sum_{i=1}^{t-1}c_{i}\left\Vert \tilde{\mathbf{p}}_{i}-\mathbf{p}\right\Vert ^{2}
\end{equation}
where $\sum_{i=1}^{t-1}c_{i}=1$. The iterative online update equation
can be obtained as 
\begin{align}
\hat{\tilde{\mathbf{p}}}_{t} & =\sum_{i=1}^{t-1}c_{i}\tilde{\mathbf{p}}_{i}\\
 & =c_{t-1}\tilde{\mathbf{p}}_{t-1}+(1-c_{t-1})\hat{\tilde{\mathbf{p}}}_{i}
\end{align}
in which trivial value $c_{i}=i^{-1}$ is employed and the corresponding
regret is given as 
\begin{align}
\frac{1}{2}\sum_{i=1}^{T}c_{i}\left[\left\Vert \tilde{\mathbf{p}}_{i}-\hat{\tilde{\mathbf{p}}}_{i}\right\Vert ^{2}-\left\Vert \tilde{\mathbf{p}}_{i}-\tilde{\mathbf{p}}^{*}\right\Vert ^{2}\right] & \leq\frac{1}{2}\sum_{i=1}^{T}c_{i}\left[\left\Vert \tilde{\mathbf{p}}_{i}-\hat{\tilde{\mathbf{p}}}_{i}\right\Vert ^{2}-\left\Vert \tilde{\mathbf{p}}_{i}-\hat{\tilde{\mathbf{p}}}_{i-1}\right\Vert ^{2}\right]\nonumber \\
 & =\frac{1}{2}\sum_{i=1}^{T}c_{i}\left\Vert \tilde{\mathbf{p}}_{i}-\hat{\tilde{\mathbf{p}}}_{i}\right\Vert ^{2}\left[1-\left(1-c_{i}\right)^{2}\right]\\
 & \leq\sum_{i=1}^{T}c_{i}^{2}\left\Vert \tilde{\mathbf{p}}_{i}-\hat{\tilde{\mathbf{p}}}_{i}\right\Vert ^{2}\leq\sum_{i=1}^{T}c_{i}^{2}\left(\|\tilde{\mathbf{p}}_{i}\|^{2}+\|\hat{\tilde{\mathbf{p}}}_{i}\|^{2}\right)\nonumber \\
 & <2N(s\log N)^{2}(2-T^{-1})
\end{align}
where for Zipf distribution $\tilde{p}_{lt}^{2}=\left(-\log\frac{l^{-s}}{\bar{N}}\right)^{2}\leq s\log N+\log\bar{N}<s\log N$
with $\bar{N}=\sum_{l}l^{-s}$. 

\emph{Remark: }ASP based OL formulation is same as PPM. Therefore,
we omit it here. Also, in simulation results, for OL schemes, the
similar results for ASP-PM are shown as PPM. 

The advantage of online learning models is the reduced number of system
parameters such as compared to online prediction methods, the number
of linear coefficients and the number of observations are not required
here. Moreover, the prediction estimates can be computed with much
lower computational complexity as compared to prediction methods in
the previous section where an optimization problem has been solved
per round. 

\section{Online Learning Models With Quasi-Time Varying \label{sec:Online-Learning-KWIK}}

In the previous section, time varying scenario was investigated where
popularity distribution was changing independently in every time slot.
Here, in this section, we consider that the coefficients of regression
change block-wise, i.e., for a given block, the content popularity
data varies as a convex auto-regressive process. For this model, we
present the online prediction based on KWIK (``know what it knows'')
framework \cite{Li2011,NIPS2007_3197}. A key aspect of KWIK model
is that the algorithm is allowed to output a special value $\emptyset$
when the prediction is tested inaccurate. The goal is to minimize
the number of invalid predictions and to maximize the accuracy of
valid predictions. Toward this, let $\mathcal{H}_{l}$ denote the
matrix whose $i^{th}$ row is given as $\mathcal{H}_{l}(i)=\left(p_{l,t-d+1},\ldots,p_{l,t}\right)=\mathbf{p}_{(l),t}^{T}$,
and similarly, $\mathcal{Y}_{l}(i)=p_{l,t+1}$. 
\begin{algorithm}
 \begin{algorithmic}[1] \Require{$\alpha_{1}$ and $\alpha_{2}$} \State{Initialize $\mathcal{H}_{l}=\{\}$ and $\mathcal{Y}_{l}=\{\}$.} \For{$t=1,2,3\ldots$} \State{observe $\mathbf{p}_{t}$ at time $t$} \For{$l=1 \ldots N$} \State{compute ${\mathbf{q}_{l}}$ and ${\mathbf{v}_{l}}$ } \If{$\|{\mathbf{q}_{l}}\|\leq\alpha_{1}$ and $\|{\mathbf{v}_{l}}\|\leq\alpha_{2}$ } \State{get the coefficients using  \[ c_{kl}=\arg\min_{\sum_{k}c_{kl}=1}\sum_{i=1}^{|\mathcal{X}_{l}|}\left\Vert \mathcal{Y}_{l}(i)-\sum_{k=1}^{d}c_{kl}\mathcal{H}_{l}(i)\right\Vert _{2}^{2} \] } \State{output $\hat{{p}}_{l,t+1}=\sum_{k=1}^{d}c_{kl}\mathbf{p}_{l,t+1-k}$} \Else{} \State{output $\emptyset$} \State{add $\left\{ {p}_{l,t-d+1},\ldots,{p}_{l,t}\right\} $ in $\mathcal{H}_{l}$ and ${p}_{l,t+1}$ in $\mathcal{Y}_{l}$. } \EndIf \EndFor \EndFor \end{algorithmic} 

\caption{KWIK learning algorithm\label{alg:KWIK-Algorithm}}
\end{algorithm}
By eigenvalue decomposition, $\mathcal{H}_{l}^{T}\mathcal{H}_{l}=\mathbf{U}_{l}\mathbf{\Lambda}_{l}\mathbf{U}_{l}^{T}$,
where $\lambda_{l1}\geq\ldots\lambda_{lk}\geq1>\lambda_{l,k+1}\geq\ldots\geq0$.
Denoting $\mathbf{U}_{lk}=\mathbf{U}(1:k)$ and $\mathbf{\Lambda}_{lk}=\text{diag}\left(\lambda_{l1},\ldots,\lambda_{lk}\right)$,
for an input $\mathbf{p}_{(l)}$, accuracy vector are given as 
\begin{align}
\mathbf{q}_{l} & =\mathcal{H}_{l}\mathbf{U}_{lk}\mathbf{\Lambda}_{lk}^{-1}\mathbf{U}_{lk}^{T}\mathbf{p}_{(l)}\\
\mathbf{v}_{l} & =\mathbf{p}^{T}\mathbf{U}_{l}(k+1:d)
\end{align}

These vectors provide a measure of uncertainty of online least squares.
When $\lambda_{lk}>1\forall k$, $\mathbf{p}_{(l)}=\mathcal{H}_{l}^{T}\mathbf{q}_{l}$,
i.e., $\mathbf{p}_{(l)}$ can be written as a linear combination of
the rows of $\mathcal{H}_{l}$. Intuitively, if $\|\mathbf{q}_{l}\|$
is small, there are many previous training samples ``similar'' to
$\mathbf{p}_{(l)}$ and hence the least squares estimate based on
$\mathcal{H}_{l}$ is likely to be accurate. For ill-conditioned $\mathcal{H}_{l}^{T}\mathcal{H}_{l}$
, $\mathbf{q}_{l}$ is undefined in this case, and the directions
corresponding to small eigenvalues should be considered as defined
in $\mathbf{v}_{l}$. 

Incorporating accuracy considerations, online KWIK process is presented
in Algorithm \ref{alg:KWIK-Algorithm}. In this algorithm, for each
file, the accuracy is checked and least square solutions are obtained. 

From \cite{NIPS2007_3197}, Algorithm \ref{alg:KWIK-Algorithm} is
an admissible algorithm with a sample complexity bound of $O(d^{3}/\epsilon^{4})$,
when $\left|\hat{p}_{l,t}-\mathbf{p}_{(l)}^{T}\mathbf{c}_{(l)}\right|\leq\epsilon$. 

\emph{Remark: }The above KWIK framework has been applied for PPM.
The similar online prediction strategy can be considered for GPM,
IPM, RPM and ASP-PM. The explicit algorithm is not shown here. However,
the simulation for each these results can be seen in the next section
as follows. 

\section{Simulation Results \label{sec:Simulation-Results}}

For simulations, we set $N=3$, $L=2$, $d=4$ and $\tau=10$. The
data for popularity profile is generated using Poisson requests model
with Zipf distribution across files with parameter $s=1.5$. This
procedure is given in Algorithm \ref{alg:Simulation-Data-Generation}.
\begin{algorithm}
\begin{algorithmic}[1] \Require{ mean requests $\lambda_{req}$, inter-arrival request time $\Delta t$, and time slot duration $T$}
\Ensure{request data matrix $N_{req}$ and popularity data $p$}
\State{Generate $t(i)\sim\exp(\Delta t)$, $n_{R}(i)\sim Poisson(\lambda_{req})$, $n_{f}(i)\sim Zipf(N,s)$, for $i=1,\ldots,N_{sim}$}
\State{Compute $t_{c}$ (cumulative sum of series $t$) } \State{set $T_{m}=\max\left\lceil t_{c}/T\right\rceil $.}
\For{$i=1,\ldots,N_{sim}$}
\If{$(i-1)T\leq t_{c}(i)<iT$}  \State{update the request data matrix \[ N_{req}\left(n_{f}(i),\left\lceil \frac{t_{c}(i)}{T}\right\rceil \right)=N_{req}\left(n_{f}(i),\left\lceil \frac{t_{c}(i)}{T}\right\rceil \right)+n_{R}(i) \]} \EndIf \EndFor
\State{Get $p(l,i)=\frac{N_{req}(l,i)}{\sum_{l}N_{req}(l,i)}$.} \end{algorithmic}

\caption{Synthetic Popularity Profile Data Generation\label{alg:Simulation-Data-Generation}}
\end{algorithm}
In this algorithm, each consumer sends $\lambda_{req}=100$ requests
in a Poisson process such that the time interval of the data requests
follows an exponential distribution with a mean of $\Delta t=0.01$
seconds.
The distribution of data requests from the
network accords with Zipf\textquoteright s law. This setup has been
simulated for $1000$ runs with $N_{sim}=2\times10^{5}$ and $T=15$
mins. PPP parameters for computing ASP are as follows: noise power
$\sigma^{2}=0$, BS density $\lambda_{BS}=200$, bandwidth $W=24$kHz,
path loss exponent $\alpha=3.5$, and rate threshold $R_{0}=1$.

\begin{figure}[t]
\centering

\includegraphics[width=8cm]{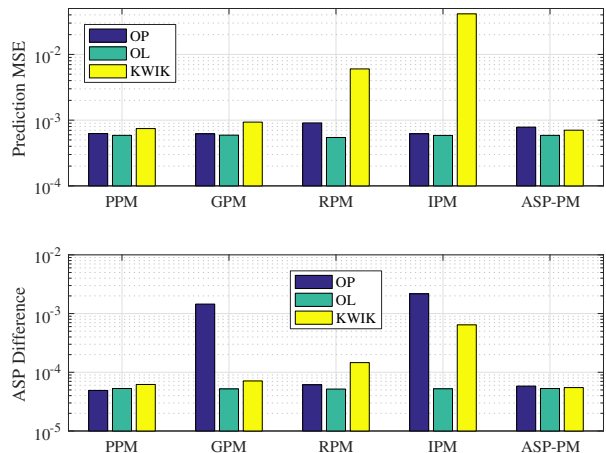}

\caption{MSE and ASP difference for different models for time varying scenario.\label{fig:Figure-illustrates-MSE}}
\end{figure}
\begin{figure}[t]
\centering

\includegraphics[width=8cm]{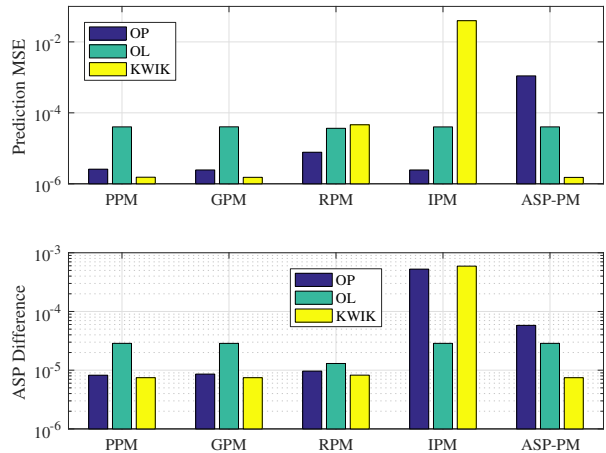}

\caption{MSE and ASP difference for different models for quasi-time varying
scenario.\label{fig:Figure-illustrates-MSE_semi}}
\end{figure}
Figure \ref{fig:Figure-illustrates-MSE} shows the MSE of popularity
prediction $(\mathbb{E}\left\Vert \mathbf{p}-\hat{\mathbf{p}}\right\Vert ^{2})$
and ASP difference $(\mathbb{E}\left[P_{s}(\mathbf{p},\mathbf{q}^{*})-P_{s}(\hat{\mathbf{p}},\hat{\mathbf{q}})\right]>0)$
achieved for different online prediction (OP) and online learning
(OL) models. It can be observed that for OP, GPM and PPM achieve the
minimum MSE, which approximates IPM, while for OL, the difference
MSE for the models is negligible. In particular, for OL, request based
method minimizes MSE. Regarding ASP difference for OP, PPM results
in minimum one, which approximates ASP based model and RPM. ASP difference
shows that the minimum MSE models do not necessarily result in optimal
ASP. Since for OL, MSE is approximately same for all models, the resultant
ASP regret is also almost similar. It can also be observed that for
time-varying case, both the MSE and ASP performances of OP and OL
methods are better than that of KWIK based models. On the other hand,
for quasi-time varying case, the KWIK methods yield better MSE as
well as ASP performance as shown in Figure \ref{fig:Figure-illustrates-MSE_semi}.
In this case, KWIK yields around 40\% improvement in MSE for PPM (OL)
and GPM (OL), while 14\% improvement in ASP difference. These KWIK
learning approaches can also be seen to significantly outperform OL
methods for quasi-time varying case

\begin{figure}[t]
\centering

\includegraphics[width=8cm]{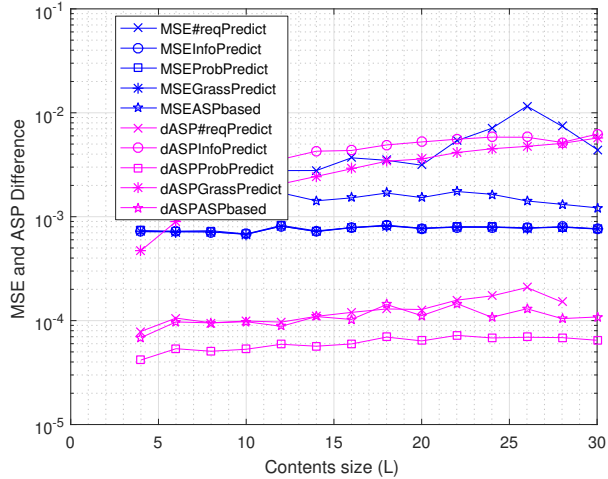}

\caption{MSE and ASP difference for different number of observations for time
varying scenario and OP models.\label{fig:Figure-depicts-OBSER}}
\end{figure}
\begin{figure}
\centering

\includegraphics[width=8cm]{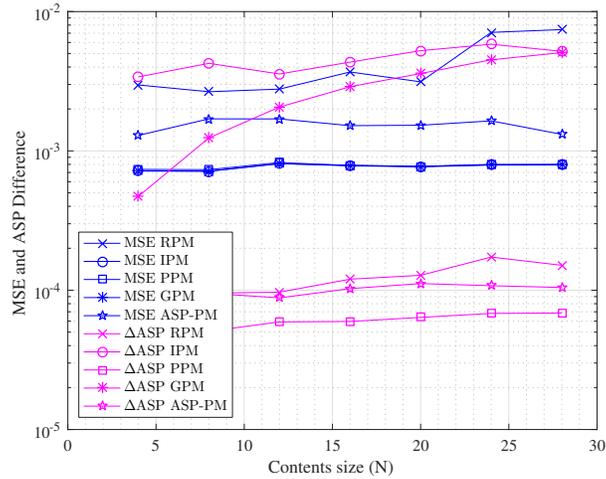}

\caption{Figure showing MSE and ASP difference for different number of contents
withwith fixed Zipf parameter $s=0.7$ and prediction duration $\tau=10$
for time varying scenario and OP models.\label{fig:Figure-depicts-L}}
\end{figure}

Figure \ref{fig:Figure-depicts-OBSER} and \ref{fig:Figure-depicts-L}
depict the variations of prediction MSE and ASP difference with respect
to number of observations and the contents size respectively for OP
models. Note that KWIK models also yield similar trend with almost
similar values hence omitted for simplicity. In the figure \ref{fig:Figure-depicts-OBSER},
it can be observed that as the number of observations increase, MSE
and ASP difference decrease for models. The order of curves for different
models is similar to Figure \ref{fig:Figure-illustrates-MSE}, i.e.,
PPM and GPM approximate prediction MSE with IPM, and PPM results into
minimum ASP difference. 
From figure \ref{fig:Figure-depicts-L}, it can be seen that as $N$
increases, MSE of prediction is approximately same except for log-request
model in which MSE increases with content size. On the other hand,
IPM and GPM show increase in ASP difference, while rest of the models
show approx similar values. IPM provides larger ASP difference than
GPM for all content lengths. 

\begin{figure}
\centering

\includegraphics[width=8cm]{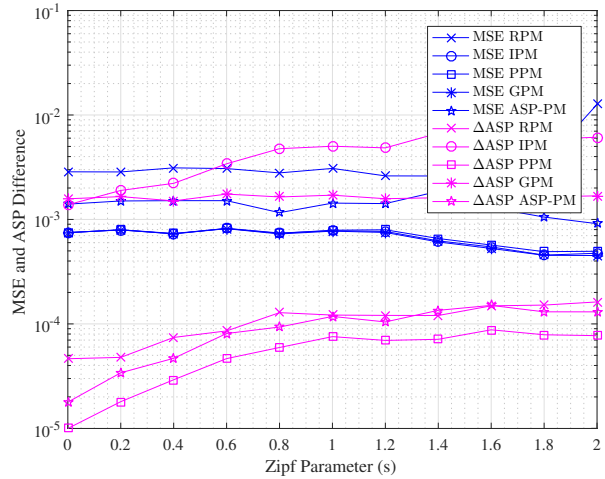}

\caption{Figure illustrates MSE and ASP difference for different Zipf parameters
with $N=10$ for time varying scenario.\label{fig:Figure-depicts-s}}
\end{figure}
Figure \ref{fig:Figure-depicts-s} presents the variations in MSE
and ASP difference of OP models as Zipf parameter is allowed to increase
with fixed $N=10$ and $\tau=10$. MSE for all models decrease slowly,
while ASP difference increases. For GPM, ASP difference is almost
same, while for other models, it increases upto a certain point and
gets constant for higher values of Zipf parameter.

\begin{figure}[t]
\centering

\includegraphics[width=8cm]{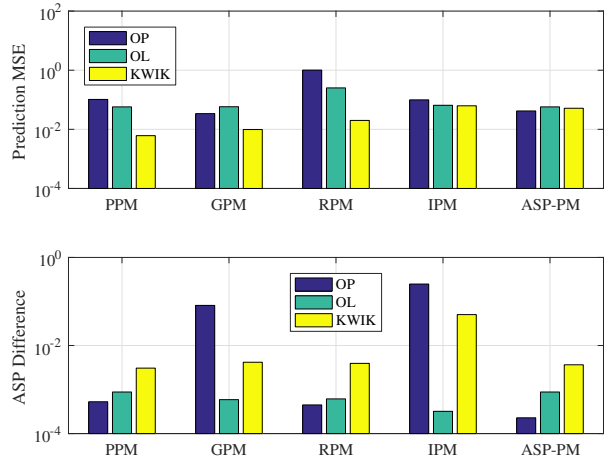}

\caption{MSE and ASP difference for different models for MovieLens dataset.\label{fig:Figure--MSE_ML}}
\end{figure}
Figure \ref{fig:Figure--MSE_ML} illustrates the prediction MSE and
ASP difference for MovieLens dataset \cite{Harper:2015:MDH:2866565.2827872}.
In this dataset, the user ratings of 100 movies are chosen with IDs
1-100 for the given timestamps. The whole duration is divided into
time slots. The popularity profile for each time slot is obtained
by normalizing the sum ratings of timestamps. For this dataset in
Figure \ref{fig:Figure--MSE_ML}, it can be observed that the prediction
MSE is minimum for PPM, followed by GPM with KWIK framework. The reason
behind is that in practice, the time varying data is correlated, for
which KWIK method has been shown in the previous results. In terms
of ASP, online prediction methods yield the better ASP. It verifies
that minimum MSE based models do not necessarily provide optimum ASP.
Among the OP methods, the better results are obtained from ASP based
model, whose construction is similar to PPM. Request base model also
gives approximately similar ASP to that of PPM. 

\section{Conclusion }\label{sec:Conclusion}

In this paper, for time varying content popularities, online prediction
and online learning models (log-request based, information based,
probability based, Grassmannian and ASP based prediction models) have
been investigated. First, for the given popularity profile, caching
probabilities have been optimized to maximize ASP of the network.
ASP difference has been derived for Grassmannian prediction model.
Also, for online learning models, the MSE regret has been bounded.
On the other hand, for quasi-time varying case, KWIK framework has
been utilized to modify the online procedure. Simulations conclude
that for time-varying case, PPM and GPM achieves minimum MSE, while
PPM results into maximum ASP, i.e., it shows that minimum MSE based
models does not necessarily result in optimal ASP. Online learning
models yields approximately similar MSE and ASP for all models than
online prediction models. Results also depicts the improvement of
KWIK framework in quasi-time varying case over other online methods,
which is verified using MovieLens dataset. 

\section*{Appendix\label{sec:Appendix}}

\subsection{ASP Derivation\label{subsec:ASP-Derivation}}

The CCDF of sum rate can be obtained as 
\begin{align}
g(p_{l}) & =\Pr\left(W\log_{2}(1+\Gamma_{il})\geq R_{0}\right)=\Pr\left(\Gamma_{il}\geq\underbrace{2^{\frac{R_{0}}{W}}-1}_{s_{0}}\right)\nonumber \\
 & =\mathbb{E}_{h_{i},r_{i},I_{i}}\Pr\left(\frac{\frac{\left|h_{i}\right|^{2}P}{r_{i}^{\alpha}}}{I_{i}+\sigma^{2}}\geq s_{0}\right)\\
 & =\mathbb{E}_{r_{i}I_{i}}\left[\mathbb{E}_{h_{i}}\Pr\left(\left|h_{i}\right|^{2}\geq s_{0}r_{i}^{\alpha}\left(\frac{I_{i}}{P}+\frac{\sigma^{2}}{P}\right)\right)\right]\\
 & =\mathbb{E}_{r_{i}}\left\{ \mathbb{E}_{I_{i}}\exp\left[-s_{0}r_{i}^{\alpha}\left(\frac{I_{i}}{P}+\frac{\sigma^{2}}{P}\right)\right]\right\} \\
 & =\mathbb{E}_{r_{i}}\exp\left[-s_{0}r_{i}^{\alpha}\left(\frac{\sigma^{2}}{P}\right)\right]\times\mathbb{E}_{h_{j},r_{j}}\exp\left[-s_{0}\sum_{j\in\Phi_{bs}\setminus\{i\}}\left|h_{j}\right|^{2}\left(\frac{r_{i}}{r_{j}}\right)^{\alpha}\right]\nonumber \\
 & =\mathbb{E}_{r_{i}}\exp\left[-s_{0}r_{i}^{\alpha}\left(\frac{\sigma^{2}}{P}\right)\right]\times\mathbb{E}_{r_{j}}\left[\prod_{j\in\Phi_{bs}(l)\setminus\{i\}}\prod_{j\in\Phi_{bs}^{c}(l)\setminus\{i\}}\frac{1}{1+s_{0}\left(\frac{r_{i}}{r_{j}}\right)^{\alpha}}\right]\label{eq:g(pl)}
\end{align}
Now, the expectation on distance will be taken \cite{schilcher2016interference}
\begin{align*}
 & \mathbb{E}_{r_{j}}\prod_{j\in\Phi_{bs}(l)\setminus\{i\}}\left(1+s_{0}r_{i}^{\alpha}r_{j}^{-\alpha}\right)^{-1}\\
 & =\exp\left(-\int_{r_{i}}^{\infty}\left(1-\frac{1}{1+s_{0}\left(\frac{r_{i}}{r_{j}}\right)^{\alpha}}\right)2\pi\lambda_{bs}p_{l}r_{j}\left(\mathrm{d}r_{j}\right)\right)\\
 & =\exp\left(-2\pi\lambda_{bs}p_{l}\int_{r_{i}}^{\infty}\left(\frac{s_{0}r_{i}^{\alpha}r_{j}^{-\alpha}}{1+s_{0}r_{i}^{\alpha}r_{j}^{-\alpha}}\right)r_{j}\left(\mathrm{d}r_{j}\right)\right)
\end{align*}
\begin{align}
 & =\exp\left(-2\pi\lambda_{bs}p_{l}\int_{r_{i}}^{\infty}\left(\frac{1}{1+\frac{r_{j}^{\alpha}}{s_{0}r_{i}^{\alpha}}}\right)r_{j}\left(\mathrm{d}r_{j}\right)\right)\\
 & =\exp\left(-\underbrace{2\pi\lambda_{bs}p_{l}s_{0}^{\frac{2}{\alpha}}r_{i}^{2}\frac{1}{\alpha}\int_{\frac{1}{s_{0}}}^{\infty}\left(\frac{u^{\frac{2}{\alpha}-1}}{1+u}\right)du}_{=Ap_{l}r_{i}^{2}}\right)\\
 & \mathbb{E}_{r_{j}}\prod_{j\in\Phi_{bs}^{c}(l)\setminus\{i\}}\left(1+s_{0}r_{i}^{\alpha}r_{j}^{-\alpha}\right)^{-1}\\
 & =\exp\left(-\int_{0}^{\infty}\left(1-\frac{1}{1+s_{0}\left(\frac{r_{i}}{r_{j}}\right)^{\alpha}}\right)\pi\lambda_{bs}(1-p_{l})\left(\mathrm{d}r_{j}^{2}\right)\right)\nonumber \\
 & =\exp\left(-\underbrace{2\pi\lambda_{bs}(1-p_{l})s_{0}^{\frac{2}{\alpha}}r_{i}^{2}\frac{1}{\alpha}\int_{0}^{\infty}\left(\frac{u^{\frac{2}{\alpha}-1}}{1+u}\right)du}_{=B\left(1-p_{l}\right)r_{i}^{2}}\right)
\end{align}
where we let $u=\frac{r_{j}^{\alpha}}{s_{0}r_{i}^{\alpha}}$, $r_{j}=\left(us_{0}\right)^{1/\alpha}r_{i}$,
$du=\alpha\frac{r_{j}^{\alpha-1}}{s_{0}r_{i}^{\alpha}}\left(dr_{j}\right)=u\alpha\frac{1}{r_{j}}\left(dr_{j}\right)$,
$r_{j}\left(\mathrm{d}r_{j}\right)=dur_{j}^{2}\frac{1}{u\alpha}=\left(us_{0}\right)^{2/\alpha}r_{i}^{2}\frac{1}{u\alpha}=u^{\frac{2}{\alpha}-1}s_{0}^{\frac{2}{\alpha}}r_{i}^{2}\frac{1}{\alpha}$.
Therefore, the resultant expression from \eqref{eq:g(pl)} can be
written as 
\begin{align}
g(p_{l}) & =\mathbb{E}_{r_{i}}\exp\left[-s_{0}r_{i}^{\alpha}\left(\frac{\sigma^{2}}{P}\right)-\left(p_{l}A+(1-p_{l})B\right)r_{i}^{2}\right]\nonumber \\
 & =\int_{0}^{\infty}\exp\left[-s_{0}r_{i}^{\alpha}\left(\frac{\sigma^{2}}{P}\right)-\left(p_{l}A+(1-p_{l})B\right)r_{i}^{2}\right]\nonumber \\
 & \qquad\times\exp(-\pi p_{l}\lambda_{bs}r_{i}^{2})2\pi p_{l}\lambda_{bs}r_{i}(dr_{i})\\
 & =p_{l}C\int_{0}^{\infty}\exp\left[-s_{0}r_{i}^{\alpha}\left(\frac{\sigma^{2}}{P}\right)\right]\\
 & \quad\times\exp\left[-\left(p_{l}A+(1-p_{l})B+p_{l}C\right)r_{i}^{2}\right](dr_{i}^{2})
\end{align}
where $C=\pi\lambda_{bs}$.

\subsection{Solution to maximization problem\label{subsec:Solution-to-maximization}}

The optimization problem inside can be written as
\begin{align}
\inf_{\mathbf{q}=[q_{1},\ldots q_{N}]} & -\sum_{i}g_{o}(q_{i})p_{i}\label{eq:inf1}\\
\text{subject to } & \sum_{i}q_{i}\leq L,\:q_{i}\in[0,1],\forall i
\end{align}
Writing the Lagrangian and KKT conditions as
\begin{align}
\mathcal{L} & =-\sum_{i}g_{o}(q_{i})p_{i}+\lambda(\sum_{i}q_{i}-L)+\sum_{i}\nu_{i}(q_{i}-1)+w_{i}q_{i}\nonumber \\
0 & =-g_{o}'(q_{i})p_{i}+\lambda+\nu_{i}-w_{i}=0\forall i\\
0 & =\lambda(\sum_{i}q_{i}-L),\lambda\geq0\forall i\\
0 & =\nu_{i}(q_{i}-1),v_{i}\geq0\forall i\\
0 & =w_{i}\left(-q_{i}\right),w_{i}\leq0\forall i
\end{align}
where $g'(q)=\frac{BC}{\left[B+q(A+C-B)\right]^{2}}$. Let $\mathcal{N}=\mathcal{Z}\cup\mathcal{P}\cup\mathcal{R}$,
where $\mathcal{Z}=\left\{ i|v_{i}=0,w_{i}<0,q_{i}=0\right\} $, $\mathcal{P}=\left\{ i|v_{i}=w_{i}=0,0<q_{i}<1\right\} $,
and $\mathcal{R}=\left\{ i|v_{i}>0,w_{i}=0,q_{i}=1\right\} .$ This
results in 
\begin{align}
q_{i} & =\left(\frac{B}{A+C-B}\right)\left[\sqrt{\frac{C}{B}\frac{p_{i}}{\lambda'_{i}}}-1\right]\\
\lambda-w_{i} & =g'(0)p_{i}=\frac{C}{B}p_{i},\forall i\in\mathcal{Z}\\
\lambda+v_{i} & =g'(1)p_{i}=\frac{C}{B}\left(\frac{B}{A+C}\right)^{2}p_{i},\forall i\in\mathcal{R}\\
\lambda & =g'(q_{i})p_{i}=\frac{C}{B}\frac{\left(\bar{\mathbf{p}}(\mathcal{P})^{T}\mathbf{1}\right)^{2}}{\eta^{2}},\forall i\in\mathcal{P}
\end{align}
where $\lambda'_{i}=\lambda+\nu_{i}-w_{i}$ and $\eta=|\mathcal{P}|+(M-\left|\mathcal{R}\right|)\frac{A-B+C}{B}$.
The objective function is simplified as 
\begin{align}
 & \sum_{i}g(q_{i})p_{i}=\sum_{i}\lambda'_{i}q_{i}\frac{B+q_{i}(A+C-B)}{B}\nonumber \\
 & =\sum_{i\in\mathcal{P}}\lambda q_{i}\sqrt{\frac{C}{B}\frac{p_{i}}{\lambda}}+\sum_{i\in\mathcal{R}}\lambda_{i}'\frac{A+C}{B}\\
 & =\left(\frac{B\lambda}{A+C-B}\right)\sum_{i\in\mathcal{P}}\left[\frac{C}{B}\frac{p_{i}}{\lambda}-\sqrt{\frac{C}{B}\frac{p_{i}}{\lambda}}\right]+\frac{C}{A+C}\sum_{i\in\mathcal{R}}p_{i}\nonumber \\
 & =\left(\frac{C}{A+C-B}\right)\left[\sum_{i\in\mathcal{P}}p_{i}-\frac{\bar{\mathbf{p}}(\mathcal{P})^{T}\mathbf{1}}{\eta}\sqrt{p_{i}}\right]+\frac{C}{A+C}\sum_{i\in\mathcal{R}}p_{i}\nonumber \\
 & =\left(\frac{C}{A+C-B}\right)\bar{\mathbf{p}}(\mathcal{P})^{T}\mathbf{Z}_{\mathcal{P}}\bar{\mathbf{p}}(\mathcal{P})+\frac{C}{A+C}\bar{\mathbf{p}}^{T}(\mathcal{R})\bar{\mathbf{p}}^{T}(\mathcal{R})\nonumber \\
 & =\left(\frac{C}{A+C-B}\right)\bar{\mathbf{p}}^{T}\bar{\mathbf{Z}}\bar{\mathbf{p}}
\end{align}
where $\bar{p}_{ik}=p_{ik}^{1/2}$, $\mathbf{Z}_{\mathcal{P}}=\mathbf{I}-\frac{\mathbf{1}\mathbf{1}^{T}}{\eta}$
and $\bar{\mathbf{Z}}=\mathcal{D}\left(\frac{A+C-B}{A+C}\mathbf{I}_{|\mathcal{R}|},\mathbf{Z}_{\mathcal{P}},\mathbf{0}_{|\mathcal{Z}|}\right)$.

\bibliographystyle{IEEEtran}
\bibliography{caching1,caching_learning}

\begin{thebibliography}{10}
\providecommand{\url}[1]{#1}
\csname url@samestyle\endcsname
\providecommand{\newblock}{\relax}
\providecommand{\bibinfo}[2]{#2}
\providecommand{\BIBentrySTDinterwordspacing}{\spaceskip=0pt\relax}
\providecommand{\BIBentryALTinterwordstretchfactor}{4}
\providecommand{\BIBentryALTinterwordspacing}{\spaceskip=\fontdimen2\font plus
\BIBentryALTinterwordstretchfactor\fontdimen3\font minus
  \fontdimen4\font\relax}
\providecommand{\BIBforeignlanguage}[2]{{%
\expandafter\ifx\csname l@#1\endcsname\relax
\typeout{** WARNING: IEEEtran.bst: No hyphenation pattern has been}%
\typeout{** loaded for the language `#1'. Using the pattern for}%
\typeout{** the default language instead.}%
\else
\language=\csname l@#1\endcsname
\fi
#2}}
\providecommand{\BIBdecl}{\relax}
\BIBdecl

\bibitem{8531745}
Y.~Jiang, M.~Ma, M.~Bennis, F.~Zheng, and X.~You, ``User preference learning
  based edge caching for fog radio access network,'' \emph{IEEE Transactions on
  Communications (Early Access)}, 2018.

\bibitem{shanmugam2013femtocaching}
K.~Shanmugam, N.~Golrezaei, A.~G. Dimakis, A.~F. Molisch, and G.~Caire,
  ``Femtocaching: Wireless content delivery through distributed caching
  helpers,'' \emph{IEEE Transactions on Information Theory}, vol.~59, no.~12,
  pp. 8402--8413, 2013.

\bibitem{poularakis2016complexity}
K.~Poularakis and L.~Tassiulas, ``On the complexity of optimal content
  placement in hierarchical caching networks,'' \emph{IEEE Transactions on
  Communications}, vol.~64, no.~5, pp. 2092--2103, 2016.

\bibitem{blaszczyszyn2015optimal}
B.~Blaszczyszyn and A.~Giovanidis, ``Optimal geographic caching in cellular
  networks,'' in \emph{IEEE International Conference on Communications (ICC)},
  2015, pp. 3358--3363.

\bibitem{serbetci2017optimal}
B.~Serbetci and J.~Goseling, ``Optimal geographical caching in heterogeneous
  cellular networks with nonhomogeneous helpers,'' \emph{arXiv preprint
  arXiv:1710.09626}, 2017.

\bibitem{liu2017caching}
D.~Liu and C.~Yang, ``Caching policy toward maximal success probability and
  area spectral efficiency of cache-enabled hetnets,'' \emph{IEEE Transactions
  on Communications}, vol.~65, no.~6, pp. 2699--2714, 2017.

\bibitem{liu2016caching}
D.~Liu, B.~Chen, C.~Yang, and A.~F. Molisch, ``Caching at the wireless edge:
  design aspects, challenges, and future directions,'' \emph{IEEE
  Communications Magazine}, vol.~54, no.~9, pp. 22--28, 2016.

\bibitem{avrachenkov2017low}
K.~Avrachenkov, J.~Goseling, and B.~Serbetci, ``A low-complexity approach to
  distributed cooperative caching with geographic constraints,''
  \emph{Proceedings of the ACM on Measurement and Analysis of Computing
  Systems}, vol.~1, no.~1, p.~27, 2017.

\bibitem{avrachenkov2017optimization}
K.~Avrachenkov, X.~Bai, and J.~Goseling, ``Optimization of caching devices with
  geometric constraints,'' \emph{Performance Evaluation}, vol. 113, pp. 68--82,
  2017.

\bibitem{maddah2014fundamental}
M.~A. Maddah-Ali and U.~Niesen, ``Fundamental limits of caching,'' \emph{IEEE
  Transactions on Information Theory}, vol.~60, no.~5, pp. 2856--2867, 2014.

\bibitem{sadeghi2018optimal}
A.~Sadeghi, F.~Sheikholeslami, and G.~B. Giannakis, ``Optimal and scalable
  caching for 5g using reinforcement learning of space-time popularities,''
  \emph{IEEE Journal of Selected Topics in Signal Processing}, vol.~12, no.~1,
  pp. 180--190, 2018.

\bibitem{yin2018prediction}
J.~Yin, L.~Li, H.~Zhang, X.~Li, A.~Gao, and Z.~Han, ``A prediction-based
  coordination caching scheme for content centric networking,'' in \emph{WOCC},
  2018, pp. 1--5.

\bibitem{liu2018content}
W.-X. Liu, J.~Zhang, Z.-W. Liang, L.-X. Peng, and J.~Cai, ``Content popularity
  prediction and caching for {ICN}: A deep learning approach with {SDN},''
  \emph{IEEE access}, vol.~6, pp. 5075--5089, 2018.

\bibitem{nakayama2015caching}
H.~Nakayama, S.~Ata, and I.~Oka, ``Caching algorithm for content-oriented
  networks using prediction of popularity of contents,'' in \emph{IFIP/IEEE
  International Symposium on Integrated Network Management (IM)}, 2015, pp.
  1171--1176.

\bibitem{zhang2018ppc}
Y.~Zhang, X.~Tan, and W.~Li, ``{PPC}: Popularity prediction caching in {ICN},''
  \emph{IEEE Communications Letters}, vol.~22, no.~1, pp. 5--8, 2018.

\bibitem{7775114}
S.~Müller, O.~Atan, M.~van~der Schaar, and A.~Klein, ``Context-aware proactive
  content caching with service differentiation in wireless networks,''
  \emph{IEEE Transactions on Wireless Communications}, vol.~16, no.~2, pp.
  1024--1036, Feb 2017.

\bibitem{7524790}
S.~Li, J.~Xu, M.~van~der Schaar, and W.~Li, ``Trend-aware video caching through
  online learning,'' \emph{IEEE Transactions on Multimedia}, vol.~18, no.~12,
  pp. 2503--2516, Dec 2016.

\bibitem{Harper:2015:MDH:2866565.2827872}
\BIBentryALTinterwordspacing
F.~M. Harper and J.~A. Konstan, ``The movielens datasets: History and
  context,'' \emph{ACM Trans. Interact. Intell. Syst.}, vol.~5, no.~4, pp.
  19:1--19:19, Dec. 2015. [Online]. Available:
  \url{http://doi.acm.org/10.1145/2827872}
\BIBentrySTDinterwordspacing

\bibitem{Boyd2010}
S.~Boyd and L.~Vandenberghe, \emph{{Convex Optimization}}, 2010, vol.~25,
  no.~3.

\bibitem{lawson1995solving}
C.~L. Lawson and R.~J. Hanson, \emph{Solving least squares problems}.\hskip 1em
  plus 0.5em minus 0.4em\relax Siam, 1995, vol.~15.

\bibitem{971250f0a1b711ddb6ae000ea68e967b}
R.~Bro and S.~Jong, ``\BIBforeignlanguage{English}{A fast non-negativity-
  constrained least squares algorithm},''
  \emph{\BIBforeignlanguage{English}{Journal of Chemometrics}}, vol.~11, pp.
  393--401, 1997.

\bibitem{el2012grassmannian}
O.~El~Ayach and R.~W. Heath, ``Grassmannian differential limited feedback for
  interference alignment,'' \emph{IEEE Transactions on Signal Processing},
  vol.~60, no.~12, pp. 6481--6494, 2012.

\bibitem{zhang2012robust}
Y.~Zhang and M.~Lei, ``Robust {Grassmannian} prediction for limited feedback
  multiuser {MIMO} systems,'' in \emph{Wireless Communications and Networking
  Conference (WCNC)}, 2012, pp. 863--867.

\bibitem{shalev2012online}
S.~Shalev-Shwartz \emph{et~al.}, ``Online learning and online convex
  optimization,'' \emph{Foundations and Trends{\textregistered} in Machine
  Learning}, vol.~4, no.~2, pp. 107--194, 2012.

\bibitem{Li2011}
L.~Li, M.~L. Littman, T.~J. Walsh, and A.~L. Strehl, ``Knows what it knows:
  a framework for self-aware learning,'' \emph{Machine Learning}, vol.~82,
  no.~3, pp. 399--443, Mar 2011.

\bibitem{NIPS2007_3197}
A.~L. Strehl and M.~L. Littman, ``Online linear regression and its application
  to model-based reinforcement learning,'' in \emph{Advances in Neural
  Information Processing Systems 20}, J.~C. Platt, D.~Koller, Y.~Singer, and
  S.~T. Roweis, Eds.\hskip 1em plus 0.5em minus 0.4em\relax Curran Associates,
  Inc., 2008, pp. 1417--1424.

\bibitem{schilcher2016interference}
U.~Schilcher, S.~Toumpis, M.~Haenggi, A.~Crismani, G.~Brandner, and
  C.~Bettstetter, ``Interference functionals in poisson networks,'' \emph{IEEE
  Transactions on Information Theory}, vol.~62, no.~1, pp. 370--383, 2016.

\end{thebibliography}

\end{document}